\documentclass[aps,pre,twocolumn,superscriptaddress,showpacs,showkeys ]{revtex4-1}

\usepackage{graphicx}                   
\usepackage{hyperref}                   
\usepackage{amsmath,amssymb}            
\usepackage{epstopdf}
\usepackage{subcaption}					

\usepackage{color}
\definecolor{orange}{rgb}{1,0.5,0}
\definecolor{brown}{rgb}{0.65, 0.16, 0.16}
\definecolor{phlox}{rgb}{0.87, 0.0, 1.0}

\graphicspath{{figs/}}					

\begin{document}

\title{Self-Avoiding Walk on the square site-diluted Ising-correlated lattice}

\author{J. Cheraghalizadeh}
\affiliation{Department of Physics, University of Mohaghegh Ardabili, P.O. Box 179, Ardabil, Iran}
\email{jafarcheraghalizadeh@gmail.com}

\author{M. N. Najafi*}
\affiliation{Department of Physics, University of Mohaghegh Ardabili, P.O. Box 179, Ardabil, Iran}
\email{morteza.nattagh@gmail.com}

\author{H. Mohammadzadeh}
\affiliation{Department of Physics, University of Mohaghegh Ardabili, P.O. Box 179, Ardabil, Iran}
\email{h.mohammadzadeh@gmail.com}

\author{A. Saber}
\affiliation{Department of Physics, University of Mohaghegh Ardabili, P.O. Box 179, Ardabil, Iran}
\email{ahad.saber@gmail.com}

\begin{abstract}
The self-avoiding walk on the square site-diluted correlated percolation lattice is considered. The Ising model is employed to realize the spatial correlations of the metric space. As a well-accepted result, the (generalized) Flory's mean field relation is tested to measure the effect of correlation. After exploring a perturbative Fokker-Planck-like equation, we apply an enriched Rosenbluth Monte Carlo method to study the problem. To be more precise, the winding angel analysis is also performed from which the diffusivity parameter of Schramm-Loewner evolution (SLE) theory ($\kappa$) is extracted. We find that at the critical Ising (host) system the exponents are in agreement with the Flory's approximation. For the off-critical Ising system we find also a new behavior for the fractal dimension of the walker trace in terms of the correlation length of the Ising system $\xi(T)$, i.e. $D_F^{\text{SAW}}(T)-D_F^{\text{SAW}}(T_c)\sim \frac{1}{\sqrt{\xi(T)}}$.
\end{abstract}


\maketitle

\section{Introduction}
The effect of environmental disorder on the critical behaviors is a long-standing problem in the condensed matter systems. Self-avoiding walk (SAW) as a realization of many physical systems (such as polymers) in such a media has attracted many attentions in the literature. The problem is more interesting when the disorder is itself critical or generally self-affine, since two diverging lengths (one for SAW and another for the host media) compete, which may lead to non-trivial effects~\cite{barat1995statistics}. Since only the end parts of SAW can lie on dead ends, the asymptotic behavior of their gyration radius is expected to be dominated by the structure of backbone, rather than by that of the full fractal lattice. On the percolation clusters the fact that the backbone fractal dimension is different from the spectral dimension of the system, reveals that random walks and SAWs probe different properties of the fractal lattice~\cite{rammal1984self}. \\
The problem of SAW in dilute systems has begun by Chakraberti \textit{et. al.}~\cite{chakrabarti1981statistics} in which based on renormalization group ideas it was shown that the properties of the model do not change by disorder which was challenged by Kremer~\cite{kremer1981self} for the dilute diamond lattice and by Aharony \textit{et. al.} for other fractal lattices~\cite{aharony1989flory} depending on their backbone fractal dimension. The findings of Kremer~\cite{kremer1981self} apparently violated the Harris criterion according to which for $\alpha>0$ ($\alpha\equiv$ the exponent of the heat capacity) systems (like SAW), the disorder should be relevant. The results showed that for $p>p_c$ ($p_c\equiv$ the percolation threshold) the $\nu$ exponent ($\equiv$ the exponent of the end-to-end distance, see following sections) is identical to the pure SAW on the regular lattice, and for $p=p_c$ it is in accordance with the Aharony generalization of the Flory's mean field formula $\nu=\frac{3}{2+\bar{d}}$ ($\approx \frac{2}{3}$ for the diamond lattice) in which $\bar{d}=d-\beta/\nu_{\text{perc}}$ and $d$ is the spatial dimension of the system and $\beta$ and $\nu_{\text{perc}}$ are the exponents of the percolation problem. It was also suggested that a more appropriate generalization of the Flory's result is $\nu=\frac{1}{\bar{d}}\left(\frac{3\tilde{d}}{2+\tilde{d}} \right) $ in which $\tilde{d}$ is the spectral dimension of the fractal space. The dependence of $\nu$ to many other fractal parameters have been investigated~\cite{rammal1984self,elezovic1987critical,dhar1988critical,milosevic1991self}. The exponent has also been derived within many other approximations~\cite{lam1984self,roy1987scaling,roy1990theory,blavatska2008scaling,blavatska2008walking}. For a good review see~\cite{barat1995statistics}. Nakayashi \textit{et. al.} argued that the large change of $\nu$ for $p=p_c$ should be the effect of large errors, and the exponent does not change even at $p=p_c$~\cite{lee1988self,lee1989monte,nakanishi1992self,rintoul1994statistics}, the result which was argued by other authors~~\cite{barat1995statistics}.\\
The most conclusive field theoretical result were found by  Meir and Harris~\cite{meir1989self} and extended later~\cite{von2004two} in which by starting from Landau-Ginsberg-Wilson Hamiltonian it was found that for non-critical disorder $p>p_c$ the RG flow is towards the pure SAW fixed point and for the critical disorder ($p=p_c$), $\nu_p=\frac{1}{2}\left( 1+\epsilon/8+15\epsilon^2/256\right)$ in which $\epsilon\equiv 4-d$. In this respect one expects that for $p>p_c$, the large spatial scales $r\gg \zeta_p$ (in which $\zeta_p$ is the percolation correlation length) the properties of pure SAW is retrieved, whereas for $r\ll \zeta_p$ the behaviors of the model at $p=p_c$ is seen~\cite{sahimi1984self}. Despite of this huge literature, the effect of the spatial correlation in the metric space in not known yet. Also the conformal invariance of the SAW on the critical fractal lattices is another important question which has not been addressed in the literature. In the present paper we introduce the correlations in the host system by means of the Ising model. The importance of these correlations is also argued in detail in terms of a Fokker-Planck-like equation. Monte Carlo method, as well as winding angel method is used to extract the various exponents of the self-avoiding random walkers, or equivalently polymers.\\
The paper has been organized as follows: In the following section, we motivate this study and introduce and describe the model. The numerical methods and details are explored in SEC~\ref{NUMDet}. The results are presented in the section~\ref{results} which contains two subsections: SEC~\ref{critical} in which the critical results are presented and SEC~\ref{offcritical} in which the power-law behaviors in the off-critical temperatures are presented. We end the paper by a conclusion.

\section{The construction of the problem}
\label{sec:model}
\begin{figure*}
	\begin{subfigure}{0.25\textwidth}\includegraphics[width=\textwidth]{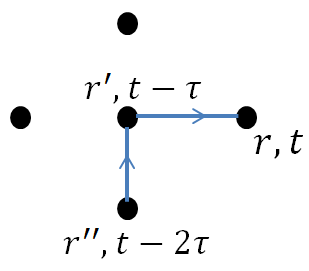}
		\caption{}
		\label{fig:situation2}
	\end{subfigure}
	\centering
	\begin{subfigure}{0.25\textwidth}\includegraphics[width=\textwidth]{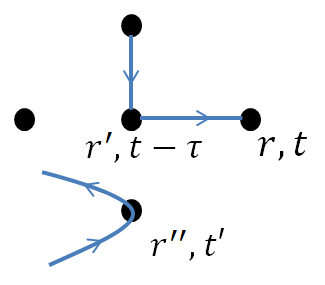}
		\caption{}
		\label{fig:situation1}
	\end{subfigure}
	\begin{subfigure}{0.25\textwidth}\includegraphics[width=\textwidth]{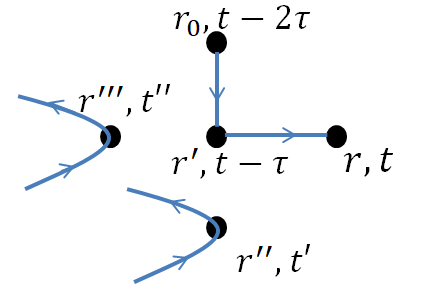}
		\caption{}
		\label{fig:situation3}
\end{subfigure}
	\caption{(Color online): Three situations of SAWs reaching the point $\vec{r}$ at time $t$.}
	\label{situations}
\end{figure*}

The notion of critical phenomena on the fractal lattices as the host was mainly begun by the work of Gefen \textit{et. al.}~\cite{gefen1980critical}. The concept can be generalized to dilute systems which become fractal in some limits. The examples are magnetic material in the porous media \cite{kose2009label,kikura2004thermal,matsuzaki2004real,philip2007enhancement,kim2008magnetic,keng2009colloidal,kikura2007cluster,najafi2016monte}, the fluid dynamics in the porous media~\cite{najafi2016water,najafi2015geometrical} and the self-organized criticality on the percolation lattices~\cite{najafi2016bak,cheraghalizadeh2017mapping}. Among the statistical models, simple random walk (RW) and self-avoiding walk (SAW) have many motivations to be investigated in dilute systems. The example is the self-non-intersecting chains of monomers in disordered systems which has strong connections to experiments~\cite{barat1995statistics}. Also correlated SAW is served as an important realization for protein folding in lattices~\cite{lau1989lattice}, in which the correlations are mostly realized by the Ising interactions between monomers~\cite{tang2000simple}. \\
In the above-mentioned literature of the SAW in disordered systems, the site-diluteness of the host media is commonly realized by the percolation theory, in which no correlation between the active sites (through which the random walker can pass) are considered, and there has been a little attention to the correlation effects in the host system. Recently it has been shown that involving the (Ising-type) correlations in the diluteness pattern of the media dramatically changes the properties of the sandpile model with respect to that on the uncorrelated percolation lattice~\cite{cheraghalizadeh2017mapping}. More interesting effects are for the critical host system in which two diverging lengths (one for the dynamical model and another for the host system) compete. This motivates one to consider the SAW on the correlated site-diluted lattices which is the aim of the present paper. The motivations of the present work is twofold:\\
1- Does the Flory's generalized relation work for SAW on the correlated dilute system? What can be said about the different exponents of SAW?\\
2- What are the behaviors of SAW in the off-critical host system? More precisely does the system exhibit power-law behaviors in this case?\\
To bring the correlations in the diluteness pattern of the host media, we have used the Ising model. In this model the correlations are simply controlled by the artificial \textit{temperature} $T$ (which has nothing to do with the real temperature) and the spins play the role of the field of activity-inactivity of the media. It is notable that the activity configuration of the media is quenched, i.e. when an Ising configuration is obtained, SAW samples are generated in the resulting dilute lattice. \\
If we show the Ising spins by $\sigma$, then $\sigma=+1$ ($\sigma=-1$) are attributed to the active (inactive) sites. By temperature we mean the control parameter which tunes the correlations of the host system. Therefore we use the Ising Hamiltonian ($H$):
\begin{equation}
H=-J\sum_{\left\langle i,j\right\rangle}\sigma_i\sigma_j-h\sum_{i}\sigma_i, \ \ \ \ \ \sigma_i=\pm 1
\label{Eq:Ising}
\end{equation}
in which $J$ is the coupling constant, $h$ is the magnetic field (which is supposed to be zero in this paper) and $\sigma_i$ and $\sigma_j$ are the spins at the sites $i$ and $j$ respectively having the values $\mp 1$ (as introduced above). $\left\langle i,j \right\rangle$ shows that the sites $i$ and $j$ are nearest neighbors. $J>0$ corresponds to ferromagnetic system (positively correlated host lattice), whereas $J<0$ is for anti-ferromagnetic ones (negatively correlated host lattice). We emphasis that in this paper we use the Ising model as the metric space and our model is not a magnetic one, instead the spins show the activity state of the sites. The artificial temperature $T$ controls the correlations and the population of the active sites to the total number of sites and also controls the heterogeneity. The population of the active site can directly be controlled by $h$ which determines the preferred direction of the spins in the Ising model. For $h=0$ the model is well-known to exhibit a non-zero magnetization per site $M=\left\langle \sigma_i\right\rangle $ at temperatures below the critical temperature $T_c$. Although we set $h=0$ throughout this paper, we prefer to mention some points concerning this parameter here. In the Ising model the magnetization has a discontinuity at $h=0$ along the $T<T_c$ line, i.e. for $h=0^+$ and $T<T_c$ we have $M>0$, whereas for the case $h=0^-$ and $T<T_c$ we have $M<0$. We can have a percolation description of the Ising model which is controlled by $T$ and $h$ as follows: In each $T$ and $h$ the system is composed of some spin clusters. Let us consider only up-spin clusters, having in mind that the system has the symmetry $h\rightarrow -h$ and $\sigma_i\rightarrow -\sigma_i$. We define $h_{th}(T)$ as the magnetic field threshold below which there is no spanning cluster of the parallel spins and above which some spanning clusters appear. Apparently for $T=0$ all spins align in the same direction and $H_{th}(T=0)=0^+$. Also for $T=\infty$ the spins are uncorrelated and take the up direction with the probability $\frac{1}{2}e^h/\cosh h$. Therefore the percolation threshold $p^{\text{Ising}}_c$ in the case $T\rightarrow\infty$ is:
\begin{equation}
p^{\text{Ising}}_c=\frac{e^{h_{th}(\infty)}}{2\cosh\left( h_{th}(\infty)\right) }.
\end{equation}
There are two transitions in the Ising model: the magnetic (paramagnetic to ferromagnetic) transition and the percolation transition (in which the connected geometrical spin clusters percolate). For the 2D regular Ising model at $h=0$ these two transitions occur simultaneously~\cite{delfino2009field}, although it is not the case for all versions of the Ising model, e.g. for the site-diluted Ising model~\cite{najafi2016monte}. \\
We define the Ising model on the $L\times L$ square lattice. Then by solving the Eq.~\ref{Eq:Ising} for $h=0$ an Ising sample at a temperature $T\leq T_c$ is made, and some self-avoiding random walkers start the motion (from the boundary of bulk, depending on the statistical observables) on the largest spanning cluster of the sample as the host media. This is defined as the connected spin cluster which contains the same spin sites and also connects opposite boundaries of the system. Note that the walks is only possible on the active ($\sigma=+1$) sites in the spanning cluster. Let us name this host area as the \textit{active space}. The active-space coordination number is defined as $z_j\equiv \sum_{i\in \delta_i}\delta_{\sigma_i,1}$
in which $\delta_i$ is the set of neighbors of $i$th site and $\delta$ is the Kronecker delta function. Let us denote the trace of a SAW up to time $t$ by $\gamma_t$ and tip of the trace by $\vec{r}_t$. Therefore when a random walker reaches the site $\vec{r}$ at time $t$, it has ways $Z_t$ to move in the next time, which is:
\begin{equation}
\begin{split}
Z_t(\vec{r}_t=\vec{r}) & \equiv z_{\vec{r}}-\sum_{t^{\prime}\leq t} \sum_{\vec{r}'\in \delta \vec{r}}\delta(\vec{r}_{t^{\prime}},\vec{r}')s_{\vec{r}'}\theta(\vec{r},\vec{r}')\\
& = \sum_{t^{\prime}\leq t}\sum_{\vec{r}'\in \delta \vec{r}} \left[\delta_{t,t^{\prime}}-\delta(\vec{r}_{t^{\prime}},\vec{r})\right]s_{\vec{r}'}\theta(\vec{r},\vec{r}')
\end{split}
\label{Z_t}
\end{equation}
In this relation $\delta \vec{r}$ is the set of active neighbors of $\vec{r}$ and $\delta(\vec{r}_t,\vec{r})=1$ if the random walker is in site $\vec{r}$ at time $t$, and zero otherwise, $s_{\vec{r}}=1$ if the site $\vec{r}$ is active and zero otherwise and $\theta(\vec{r}',\vec{r})\equiv 1-\delta(\vec{r}'',\vec{r})$ which is apparently unity in this expression, but becomes important in the following perturbation expansions. Note that $s_{\vec{r}}=\frac{1}{2}\left(\sigma(\vec{r})+1\right)$. When $\sum_{t^{\prime}\leq t} \sum_{\vec{r}'\in \delta \vec{r}}\delta(\vec{r}_{t^{\prime}},\vec{r}')s_{\vec{r}'}= z_{\vec{r}}$, there will be no way to move further and a new process should start. Otherwise each of the $Z_t(\vec{r}_t)$ sites is chosen with the same probability, i.e. $1/Z_t(\vec{r}_t)$. \\
In the other hand by defining $p(\vec{r},t)\equiv \left\langle \delta(\vec{r}_t,\vec{r})\right\rangle $ (in which $\left\langle \right\rangle $ is the ensemble average \textit{for fixed disorder configuration}. For the averaging over both random walks and disorder we should use the notation $\ll \gg$), one can easily show that:
\begin{equation}
\begin{split}
&p(\vec{r},t)=s_{\vec{r}} \left\langle \Theta(\vec{r},t-\tau)\sum_{\vec{r}^{\prime}\in \delta \vec{r}} \frac{\delta(\vec{r}_{t-\tau},\vec{r}^{\prime})}{Z_{t-1}(\vec{r}_{t-\tau}=\vec{r}^{\prime})}s_{\vec{r}'}\right\rangle \\
&=\sum_{n=0}^{\infty}\sum_{\vec{r}^{\prime}\in \delta \vec{r}}s_{\vec{r}}s_{\vec{r}'}\left\langle \frac{\Theta(\vec{r},t-\tau)\delta(\vec{r}_{t-\tau},\vec{r}^{\prime})}{z_{\vec{r}^{\prime}}}\epsilon(t,\vec{r}^{\prime})^n \right\rangle
\end{split}
\label{Eq:p}
\end{equation}
in which $\epsilon(t,\vec{r}')\equiv \frac{\sum_{t^{\prime}\leq t-\tau} \sum_{\vec{r}''\in \delta \vec{r}'}\delta (\vec{r}_{t^{\prime}},\vec{r}'')\theta(\vec{r}'',\vec{r})s_{\vec{r}''}}{z_{\vec{r}'}}< 1$ and $\Theta(\vec{r},t)\equiv 1-\sum_{t^{\prime}\leq t} \delta(\vec{r}_{t^{\prime}},\vec{r})$ is a non-local detector operator which is unity if the point $\vec{r}$ has not been visited up to time $t$ and zero otherwise (i. e. $w(\vec{r},t)\equiv 1-P(\vec{r},t)\equiv\left\langle \Theta(\vec{r},t)\right\rangle = 1-\sum_{t^{\prime}\leq t} p(\vec{r},t)$). The factor $s_{\vec{r}}\times\Theta(\vec{r},t-\tau)$ in the right hand of the first line of Eq.~\ref{Eq:p} assures that the site $\vec{r}$ is active and has not been visited before. It is notable that this equation is true for a quenched percolation configuration of the metric system, i.e. $\left\lbrace s_i\right\rbrace_{i=1}^N$, and has not been averaged over $s_i$'s.\\
Ignoring $n\geq 3$ terms, we reach to the relation:
\begin{equation}
\begin{split}
&\partial_t p(\vec{r},t)-Dw(\vec{r},t)\nabla^{\alpha}p(\vec{r},t) =\\
& \sum_{\vec{r}'\in \delta \vec{r}}\sum_{\vec{r}''\in \delta \vec{r}'}\frac{s_{\vec{r}}s_{\vec{r}'}s_{\vec{r}''}}{z(\vec{r}')^2}I_1(\vec{r},\vec{r}',\vec{r}'',t-\tau)\\
&+\sum_{\begin{array}{c} {\scriptstyle \vec{r}'\in \delta \vec{r}} \\ {\scriptstyle \vec{r}'',\vec{r}'''\in \delta \vec{r}'} \end{array}}\frac{s_{\vec{r}}s_{\vec{r}'}s_{\vec{r}''}s_{\vec{r}'''}}{ z(\vec{r}')^3} I_2(\vec{r},\vec{r}',\vec{r}'',\vec{r}''',t-\tau)\\
&+O(\epsilon^3)
\end{split}
\end{equation}
in which $\partial_t p(\vec{r},t)\equiv\frac{1}{\tau}\left( p(\vec{r},t)-p(\vec{r},t-\tau)\right)$, $\nabla^{\alpha}p(\vec{r},t)\equiv a^{-\alpha}\left( \sum_{\vec{r}'\in \delta \vec{r}}s_{\vec{r}}s_{\vec{r}'}\frac{p(\vec{r}',t-\tau)}{z_{\vec{r}'/4}}-4p(\vec{r},t-\tau)\right)$, $D\equiv \frac{a^{\alpha}}{4\tau}$, $a$ is the lattice constant and $\alpha$ is the order of fractional derivative, which is not \textit{a priori} known and should be determined by the fractal dimension of the host. It is evident that $\alpha=2$ for a regular lattice. It is notable that $\delta(\vec{r}_t,\vec{r})\Theta(\vec{r},t)\equiv \delta(\vec{r}_t,\vec{r})$. In the above equation we have defined:
\begin{equation}
\begin{split}
& I_1(\vec{r},\vec{r}',\vec{r}'',t)\equiv \theta(\vec{r}'',\vec{r})\frac{1}{\tau}\int_0^{t}\left\langle \Theta(\vec{r},t)\delta (\vec{r}',t)\delta (\vec{r}'',t')\right\rangle dt'\\
& I_2(\vec{r},\vec{r}',\vec{r}'',\vec{r}''',t) \equiv \theta(\vec{r}'',\vec{r})\theta(\vec{r}''',\vec{r})\times\\
&\frac{1}{\tau}\iint_0^{t}\left\langle \Theta(\vec{r},t)\delta (\vec{r}',t)\delta (\vec{r}'',t')\delta (\vec{r}''',t'')\right\rangle dt'dt''
\end{split}
\end{equation}
To find $I_1$ we write it as the form:
\begin{equation}
\begin{split}
I_1=&\frac{1}{\tau}\int_0^{t-\tau}\left\langle \Theta(\vec{r},t-\tau)\delta (\vec{r}',t-\tau)\delta (\vec{r}'',t')\right\rangle \theta(\vec{r}'',\vec{r})dt'\\
&\times (1-\delta(t',t-2\tau)+\delta(t',t-2\tau))\\
& = \frac{1}{\tau}\left\langle \Theta(\vec{r},t-\tau) \delta (\vec{r}',t-\tau)\delta (\vec{r}'',t-2\tau)\right\rangle\theta(\vec{r}'',\vec{r})\\
&+\frac{1}{\tau}\int_0^{t-\tau} \left\langle \Theta(\vec{r},t-\tau)\delta (\vec{r}',t-\tau)\delta (\vec{r}'',t')\right\rangle \\
& \times(1-\delta(t',t-2\tau))\theta(\vec{r}'',\vec{r})dt'
\end{split}
\end{equation}
For the first case we have the situation shown in Fig.\ref{fig:situation2} and the second line is equivalent to the Fig.~\ref{fig:situation1}. We have separated these two terms, since their behaviors are expected to be different. The quantities $\delta (\vec{r}',t-\tau)$ and $\delta (\vec{r}'',t-2\tau)$ in $\left\langle \delta (\vec{r}',t-\tau)\delta (\vec{r}'',t-2\tau)\right\rangle$ are maximally correlated and their multiplication forms a new field $\delta^{(2)}(\vec{r},\vec{r}',\vec{r}'',t)\equiv \frac{1}{\tau}\Theta(\vec{r},t)\delta (\vec{r}',t-\tau)\delta (\vec{r}'',t-2\tau)$ and $p^{(2)}(\vec{r}',\vec{r}'',t)\equiv \left\langle \delta^{(2)}(\vec{r}',\vec{r}'',t)\right\rangle $. For the second term however the fields are expected to be nearly independent, due to their temporal distance, i.e. $\int_0^{t-\tau} \left\langle \Theta(\vec{r},t)\delta (\vec{r}',t-\tau)\delta (\vec{r}'',t')\right\rangle (1-\delta(t',t-2\tau))dt'\approx  w(\vec{r},t-\tau)p(\vec{r}',t-\tau)P(\vec{r}'',t-2\tau)$. Therefore
\begin{equation}
\begin{split}
I_1 & \simeq \theta(\vec{r}'',\vec{r})\left[ p^{(2)}(\vec{r}',\vec{r}'',t)\right.\\ 
& \left. +\frac{1}{\tau}w(\vec{r},t-\tau)p(\vec{r}',t-\tau)P(\vec{r}'',t-2\tau)\right] .
\end{split}
\end{equation}
Now let us consider the second integral. By multiplying the following expression (which is equal to unity) :
\begin{equation}
\begin{split}
&1=\left[  (1-\delta(t',t-2\tau))(1-\delta(t'',t-2\tau))+\delta(t',t-2\tau)\right. \\
&\left. +\delta(t'',t-2\tau)-\delta(t',t-2\tau)\delta(t'',t-2\tau)\right] 
\end{split}
\end{equation}
to $I_2$ we can do the same procedure. The contribution of the first term of the second line has been shown in Fig.~\ref{fig:situation3}, and the other terms only improve the contributions of Figs~\ref{fig:situation1} and~\ref{fig:situation2}. If we pick up only the first term, we obtain :
\\
\begin{equation}
\begin{split}
&\tau I_2(\text{first term})=\theta(\vec{r}'',\vec{r})\theta(\vec{r}''',\vec{r})\times\\
&\iint_0^{t-\tau}\left\langle \Theta(\vec{r},t)\delta (\vec{r}',t-\tau)\delta (\vec{r}'',t')\delta (\vec{r}''',t'')\delta (\vec{r}_0,t-2\tau)\right\rangle \\
&\times(1-\delta(t',t-2\tau))(1-\delta(t'',t-2\tau))dt'dt''\\
&\approx \tau \theta(\vec{r}'',\vec{r})\theta(\vec{r}''',\vec{r})p^{(2)}(\vec{r}',\vec{r}_0,t)P(\vec{r}'',t-\tau)P(\vec{r}''',t-\tau)
\end{split}
\end{equation}
In this equation we have inserted a trivial term $\delta (\vec{r}_0,t-2\tau)$ in the expression in which $\vec{r}_0$ has been shown in Fig.~\ref{fig:situation3}. This insertion change nothing, since the random walker has apparently been in $\vec{r}_0$ at time $t-2\tau$. We reach finally to the relation:
\begin{equation}
\begin{split}
&\partial_t p(\vec{r},t)-Dw(\vec{r},t)\nabla^{\alpha}p(\vec{r},t) =\\
& \sum_{\vec{r}'\in \delta \vec{r}}\sum_{\vec{r}''\in \delta \vec{r}'}\frac{s_{\vec{r}}s_{\vec{r}'}s_{\vec{r}''}}{z(\vec{r}')^2}\theta(\vec{r}'',\vec{r})\left[ p^{(2)}(\vec{r}',\vec{r}'',t)\right.\\ 
& \left. +\frac{1}{\tau}w(\vec{r},t-\tau)p(\vec{r}',t-\tau)P(\vec{r}'',t-2\tau)\right]\\
&+\sum_{\begin{array}{c} {\scriptstyle \vec{r}'\in \delta \vec{r}} \\ {\scriptstyle \vec{r}'',\vec{r}'''\in \delta \vec{r}'} \end{array}}\frac{s_{\vec{r}}s_{\vec{r}'}s_{\vec{r}''}s_{\vec{r}'''}}{ z(\vec{r}')^3} \theta(\vec{r}'',\vec{r})\theta(\vec{r}''',\vec{r})\times\\
&p^{(2)}(\vec{r}',\vec{r}_0,t)P(\vec{r}'',t-\tau)P(\vec{r}''',t-\tau)\\
&+O(\epsilon^3)
\end{split}
\label{mainEQ}
\end{equation}
In this equation $P(\vec{r},t)$ needs full information of the status of the walker in the past times, i.e. it is a field which carries the information of the history of the random walk, whereas $p^{(2)}$ is a local field. Now let us average over the disorder, i.e. take configurational average from the Eq.~\ref{mainEQ} over the $s_i$ configuration. The correlation of the noise $\left\lbrace s_i\right\rbrace_{i=1}^N$ surely affect the resultant equation. The equation involves the moments of $s$ (up to fourth moment in the above equation). When $s_i$'s are uncorrelated, we obtain that $\left\langle s_{\vec{r}_1}s_{\vec{r}_2}...s_{\vec{r}_n}\right\rangle$ are equal to $s^n$ in which $s=\left\langle s_{\vec{r}}\right\rangle $. Also $\left\langle s_{\vec{r}_1}s_{\vec{r}_2}...s_{\vec{r}_n}G\right\rangle $ (in which $G$ is some function which depends on the $s$ configuration) can safely approximated by $s^n\left\langle G\right\rangle $. Apparently this is not true for correlated noises in which $\left\langle s_{\vec{r}_1}s_{\vec{r}_2}\right\rangle \neq s^2$. In the above equation $\Gamma^{(2)}(|\vec{r}-\vec{r}'|)\equiv \left\langle s_{\vec{r}}s_{\vec{r}'} \right\rangle $, $\Gamma^{(3)}(\vec{r}_1,\vec{r}_2,\vec{r}_3)\equiv \left\langle s_{\vec{r}_1}s_{\vec{r}_2}s_{\vec{r}_3} \right\rangle $ and $\Gamma^{(4)}(\vec{r}_1,\vec{r}_2,\vec{r}_3,\vec{r}_4)\equiv \left\langle s_{\vec{r}_1}s_{\vec{r}_2}s_{\vec{r}_3}s_{\vec{r}_4} \right\rangle $ have appeared which are temperature dependent and are calculated using the Ising autocorrelation functions. \\
This analysis has been presented to highlight the important effect of correlations in the metric space of the SAW. The fact that the obtained equation is perturbative and involves non-linear-non-local functions make it less efficient and therefore numerical studies are crucial in understanding its properties. 
\section{Numerical methods, Monte Carlo approach and SLE theory}\label{NUMDet}
In this paper we have used the enriched Rosenbluth method. To describe the method, let us consider a growing polymer chain (or a self-avoiding random walker) at the $n$th step, for which $(n+1)$th monomer should be added to the chain in an active neighboring site. In the Rosenbluth-Rosenbluth (RR) method we give weight $W(N) \equiv\left(\prod_{t=1}^{N}Z_t\right)^{-1}$ to each sample configuration, in which $Z_t$ has been defined in \ref{Z_t}. The configurational average is then defined by:
\begin{equation}
\left\langle R^2\right\rangle\equiv \frac{\sum_i W_i(N)R_i^2}{\sum_i W_i(N)}
\end{equation}
in which $i$ runs over distinct configurations and $W_i$ is its weight and $R_i$ is the end-to-end distance. It is notable that all polymers have the same length $N$ in this averaging. The enrichment procedure is as follows: If $W(N)$ is above a certain threshold, we add a new walker and give the new and old walker half the original weight. If $W(N)$ is below a certain threshold, then we eliminate it with the probability $p=1/2$ and double the weights of the remaining half.

By means of this method, we calculate the $\nu$ exponent (of the end-to-end distance, to be defined later) as well as the fractal dimension of SAW using the box-counting method. To be more precise, we have also used the winding-angel statistics to extract the diffusivity parameter of Schramm-Loewner evolution (SLE).\\
As a well-known fact, the critical 2D models have special algebraic and geometrical properties. The algebraic properties of these models are described within the conformal field theories. However these theories are unable to uncover the geometrical features of these models since it concerns the local fields defined in these models. SLE theory aims to describe the interfaces of two-dimensional (2D) critical models via growth processes. Thanks to this theory, a deep connection between the local properties and the global (geometrical) features of the 2D critical models has been discovered. These non-intersecting interfaces are assumed to have two essential properties, conformal invariance and the domain Markov property~\cite{cardy2005sle}.\\
In the SLE theory one replaces the critical curve by a dynamical one. We consider the model on the upper half plane, i.e. $H=\left\lbrace z\in \mathbb{C}, \Im z\geq 0\right\rbrace$. Let us denote the curve up to time $t$ as $\gamma_t$ and the hull $K_t$ as the set of points which are located exactly on the $\gamma_t$ trace, or are disconnected from the infinity by $\gamma_t$. The complement of $K_t$ is $H_{t}:=H\backslash{K_{t}}$ which is simply-connected. According to Riemann mapping theorem there is always a conformal mapping $g_t(z)$ (in two dimensions) which maps $H_{t}\rightarrow{H}$. The map $g_{t}(z)$ (commonly named as uniformizing map, meaning that it uniformizes the $\gamma_t$ trace to the real axis) is the unique conformal map with $g_{t}(z)=z+\frac{2t}{z}+O(\frac{1}{z^{2}})$ as $z\rightarrow{\infty}$ known as hydrodynamical normalization. Loewner showed that this mapping satisfies the following equation~\cite{Schramm2000Scaling,cardy2005sle,lowner1923untersuchungen,Smirnov2007Conformal}:
\begin{equation}
\partial_{t}g_{t}(z)=\frac{2}{g_{t}(z)-\xi_{t}},
\label{Loewner}
\end{equation}
with the initial condition $g_{t}(z)=z$ and for which the tip of the curve (up to time $t$) is mapped to the point $\xi_t$ on the real axis. For fixed $z$, $g_{t}(z)$ is well-defined up to time $\tau_{z}$ for which $g_{t}(z)=\xi_{t}$. The more formal definition of hull is therefore $K_{t}=\overline{\lbrace z\in H:\tau_{z}\leq t \rbrace}$. For more information see \cite{cardy2005sle,lowner1923untersuchungen}. For the critical models, it has been shown~\cite{Schramm2000Scaling} that $\xi_{t}$ (referred to as the driving function) is a real-valued function proportional to the one-dimensional Brownian motion $\xi_{t}=\sqrt{\kappa}B_{t}$ in which $\kappa$ is known as the diffusivity parameter. SLE aims to analyze these critical curves by classifying them to the one-parameter classes represented by $\kappa$. The relation between the fractal dimension of the curves $D_f\equiv \frac{1}{\nu}$ and the diffusivity parameter ($\kappa$) is $D_f=1+\frac{\kappa}{8}$. \\
The important tests of SLE are left passage probability~\cite{najafi2013left,Najafi2015Fokker}, direct SLE mapping~\cite{Najafi2012Observation,najafi2015observation} and the winding angel statistics~\cite{duplantier1988winding}. The latter is defined by the relation:
\begin{equation}
\left\langle \theta^2\right\rangle = \kappa\log R
\end{equation}
in which $R$ is end-to-end distance and $\theta$ is the total winding angel of the movement at the end point and a global direction. Noting that $R\sim l^{\nu}$, one finds that $\left\langle \theta^2\right\rangle = 8\nu(D_F-1)\log l$. This slope is exactly $2$ for the SAW on the regular lattice, i.e. $T=0$.\\
In the box-counting scheme, it is defined by the relation $N(L)\sim L^{D_f}$ in which $l$ is the length of the stochastic curve (SAW) inside a box of linear size $L$.

\subsection{Numerical details}
At $T=T_c$ for which the Ising model becomes critical, some power-law behaviors emerge. The method to simulate the system in the vicinity of this point is important, due to the problem of critical slowing down. To avoid this problem we have used the Wolff Monte Carlo method to generate Ising samples. Our ensemble averaging contains both random walks averaging as well as Ising-percolation lattice averaging. For the latter case we have generated $2\times 10^3$ Ising uncorrelated samples for each temperature on the lattice size $L=2048$. To make the Ising samples uncorrelated, between each successive sampling, we have implied $L^2/3$ random spin flips and let the sample to equilibrate by $500L^2$ Monte Carlo steps. The main lattice has been chosen to be square, for which the Ising critical temperature is $T_c\approx 2.269$. Only the samples with temperatures $T\leq T_c$ have been generated, since the spanning clusters (active space) are present only for this case. As stated in the previous section the random walkers move only on the active space which is defined as the set of connected sites having the same (majority) spin which connects two opposite boundaries. The temperatures considered in this paper are $T=T_c-\delta t_1\times i$ ($i=1,2,...,5$ and $\delta t_1=0.01$) to obtain the statistics in the close vicinity of the critical temperature $T_c\simeq 2.269$ (note that the model shows non-trivial power-law behaviors in the vicinity of the critical temperature) and $T=T_c-\delta t_2\times i$ ($i=1,2,...,10$ and $\delta t_2=0.05$) for the more distant temperatures. To equilibrate the Ising sample and obtain the desired samples we have started from the high temperatures ($T>T_c$). For each temperature $2\times 10^6$ SAWs were generated for $2\times 10^3$ Ising samples (for each Ising sample $10^3$ avalanche samples were generated and each Ising sample had its own particle dynamics to reach a steady state). We have used the Hoshen-Kopelman~\cite{hoshen1976percolation} algorithm for identifying the clusters in the lattice.\\
Once a spanning Ising percolation cluster is obtained, the SAW simulations begin. Figure \ref{fig:n=2000,512} is a $512\times 512$ sample at $T=T_c$ in which the red (white) sites show the inactive (active) sites and a $N=2000$ length SAW (which has started from the bulk and moves only on the white sites) has been shown in blue. The geometrical properties of these walks is investigated in this paper. \\
\begin{figure}
	\centerline{\includegraphics[scale=.2]{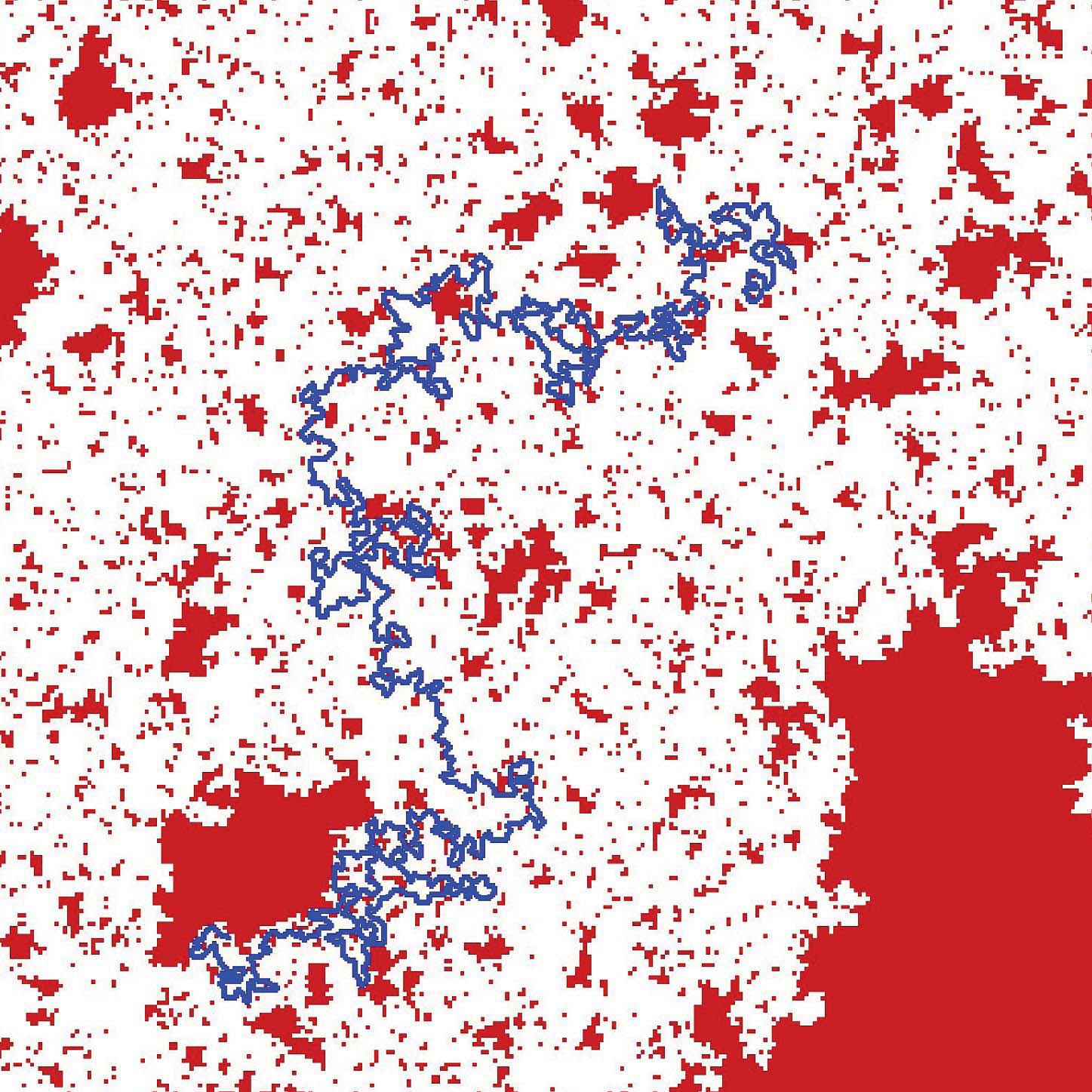}}
	\caption{A $N=2000$ bulk SAW sample in an Ising sample media in a $512\times 512$ lattice at $T=T_c$ (blue lines). The red sites represent the forbidden (inactive) sites and the white sites are representative of the active ones.}
	\label{fig:n=2000,512}
\end{figure}

\section{Results}\label{results}

Two cases have been considered separately: The critical fractal (self-affine) host space ($T=T_c$) and the super-critical one $T<T_c$. In the latter case the trend of the exponents to the critical case is obtained. It is expected that the critical behaviors of the ordinary SAW on the regular lattice is retrieved in the limit $T\rightarrow 0$. For all temperatures in the range $T<T_c$ the power-law behaviors have been observed. We argue that there are two fixed points in the problem, namely $T=0$ (the IR fixed point) and $T=T_c$ (the UV fixed point). The critical exponents are nearly constant for most parts of the phase space and show deviations in the vicinity of the critical temperature.

\subsection{Critical Temperature}\label{critical}

\begin{figure*}
	\centering
	\begin{subfigure}{0.49\textwidth}\includegraphics[width=\textwidth]{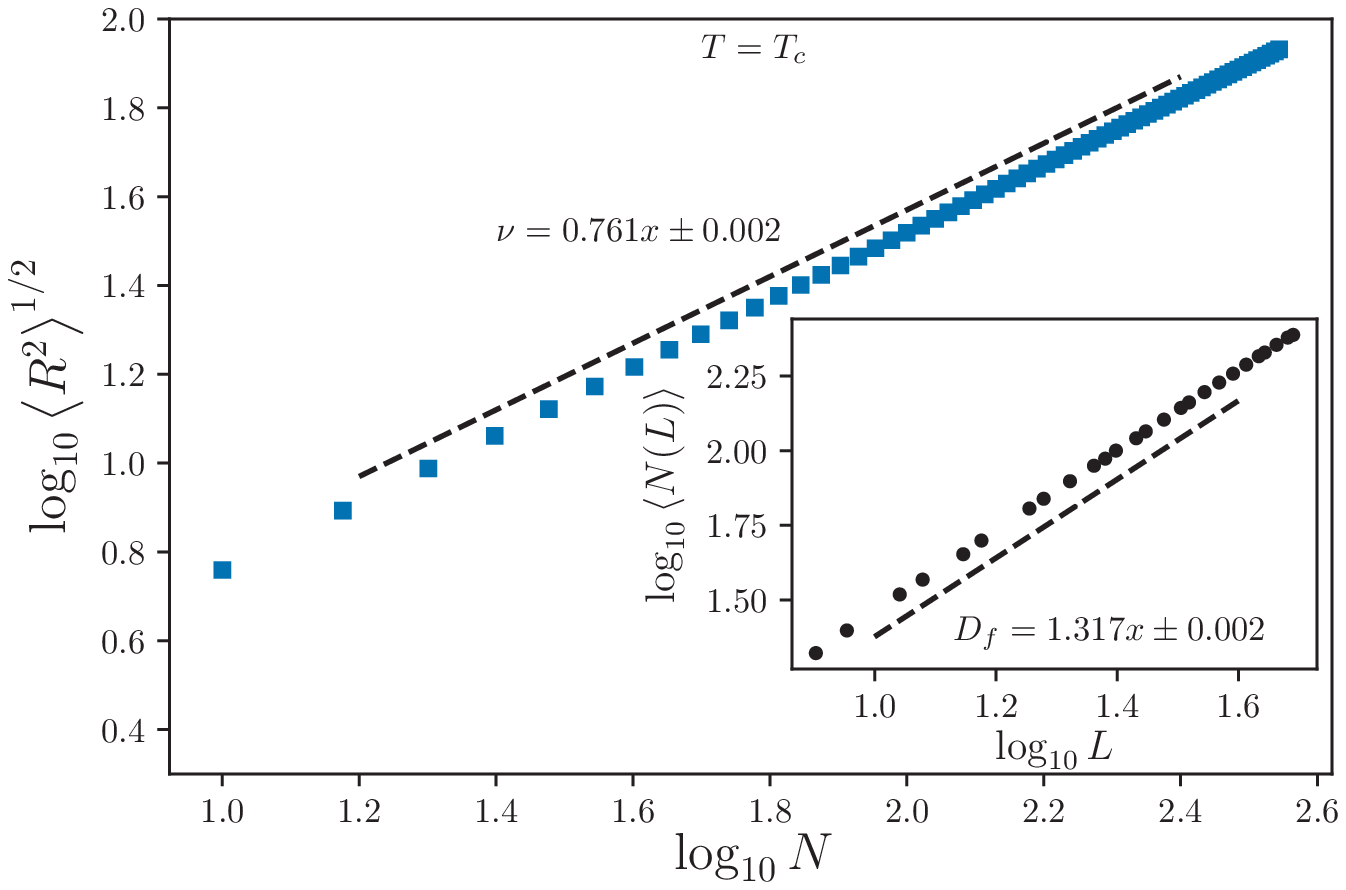}
		\caption{}
		\label{fig:r_Tc}
	\end{subfigure}
	\begin{subfigure}{0.49\textwidth}\includegraphics[width=\textwidth]{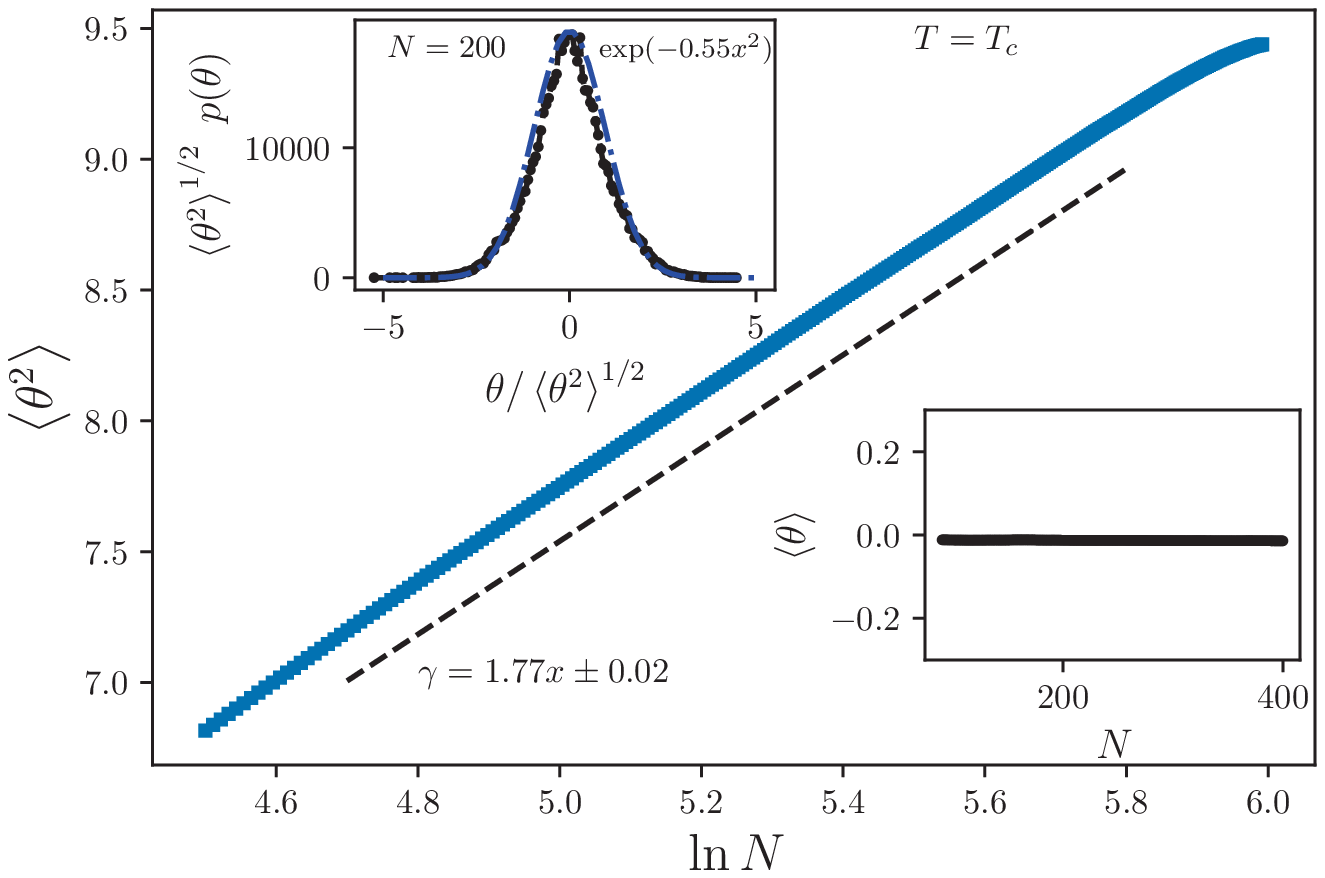}
		\caption{}
		\label{fig:WA-Tc}
	\end{subfigure}
	\caption{(Color online): (a)  $\log_{10}\left\langle R^2\right\rangle^{\frac{1}{2}}$ in terms of $\log_{10}N$ for $T=T_c$. Inset: $\log_{10}\left\langle N(L)\right\rangle$ in terms of $\log_{10}L$. (b) $\left\langle \theta^2\right\rangle $ in terms of $\ln N$ with the slope $\gamma=1.77\pm 0.02$ for $T=T_c$. Upper inset: the distribution function of $\theta$ for $N=200$. Lower inset: $\left\langle \theta\right\rangle$ in terms of $N$ which is zero.}
	\label{fig:Tc}
\end{figure*}

The characterization of the fixed points in any perturbed statistical model is very important, since it yields information about its large scale behaviors. Some critical systems on the un-correlated percolation lattice show a fixed point at $p=p_c$ (which is unstable towards the stable $p=1$ fixed point)~\cite{najafi2016water,najafi2016bak}. The Ising metric space when is seen as a percolation lattice has a chance to show the similar phenomenon, i.e. has a fixed point at $T=T_c$.
In this section we concentrate on the critical temperature case $T=T_c$. The run times in this case is large due to critical slowing down. \\

The Fig.\ref{fig:r_Tc} shows $\log\sqrt{\left\langle R^2\right\rangle}$ in terms of $\log N$ (the length of polymer) which is linear with the well-defined slope $\nu=0.761\pm 0.002$. This corresponds to $D_F=1.314\pm 0.003$. This is confirmed by the inset graph in which $\log N(L)$ has been sketched in terms of $\log L$ with the exponent $D_F^{\text{box counting}}=1.317\pm 0.002$. Therefore the Flory's relation predicts that the effective dimension of the critical Ising model is $\bar{d}^{\text{Flory}}=1.94$. This should be compared with the obtained fractal dimension of the critical Ising model which is $\bar{d}=\frac{187}{96}\simeq 1.948$ for which the Flory's approximation yields $\nu_{T_c}^{\text{Flory}}=\frac{288}{379}\simeq 0.760$~\cite{duplantier1989exact}. It is also notable that the exponents of the SAW on the critical uncorrelated percolation clusters are $\bar{d}_{p_c}^{(\text{2D percolation})}\frac{91}{49}\simeq 1.857$, $\nu_{p_c}^{(\text{SAW on 2D percolation})}=\frac{147}{189}\simeq 0.77$ and $D_F^{(\text{SAW on 2D percolation})}\simeq 1.286$. The excellent agreement between our Monte Carlo calculations and the Flory's approximation encourages one to extend this theory to all temperatures. \\
The SLE diffusivity parameter is consistent with Flory's approximation. To study this, the winding angel test has been calculated (Fig~\ref{fig:WA-Tc}). It is seen that $\left\langle \theta^2\right\rangle $ behaves logarithmically with respect to $l$ which confirms the prediction of the SLE theory. The lower inset reveals that $\left\langle \theta \right\rangle =0$ and the upper inset shows the Gaussian form of $p(\theta)$. The slope of the semi-log plot is $1.76\pm 0.04$ which is equivalent to diffusivity parameter $\kappa=2.26\pm 0.07$. Therefore the universality class of SAW$_{T=T_c}$ is distinct from the one for SAW on the regular lattice for which $\kappa=\frac{8}{3}\simeq 2.67$, i.e. $\delta\kappa\equiv \kappa_{T=0}^{\text{SAW}}-\kappa_{T=T_c}^{\text{SAW}}=0.41\pm 0.07$. The other exponents corresponding to winding angel test are $\nu^{(\kappa)}=0.778\pm 0.005$, $D_F^{\kappa}=1.284\pm 0.008$ which are more or less in agreement with the above results. The results have been gathered in TABLE~\ref{tab:nu}.
\begin{table}
	\begin{tabular}{c | c c c c c c}
		\hline & $\kappa$ & $\nu$ & $D_F$ & $\bar{d}^{\text{Flory}}$ & $c_{\kappa}$ & $t$ \\
		\hline direct & $2.51(3)$ & $0.761(2)$ & $1.314(3)$ & $1.94(1)$ & $-0.32(1)$ & $1.68(1)$\\
		\hline SLE & $2.27(6)$ & $0.778(5)$ & $1.284(8)$ & $1.85(2)$ & $-0.96(4)$ & $1.32$\\
		\hline
	\end{tabular}
	\caption{The exponents of SAW$_{T=T_c}$ obtained by two methods. In th first raw we have reported the results for the direct calculating $\nu$ and $D_F$, whereas the second raw has been obtained by calculating $\kappa$, using the winding angel distribution. The central charge has been calculated using the relation $c=1-\frac{(6-\kappa)(3\kappa-8)}{2\kappa}$ and the $t$ parameter has been defined by the relation $c=1-\frac{6}{t(t+1)}$.}
	\label{tab:nu}
\end{table}

\subsection{Off-Critical Temperatures}\label{offcritical}

\begin{figure*}
	\centering
	\begin{subfigure}{0.49\textwidth}\includegraphics[width=\textwidth]{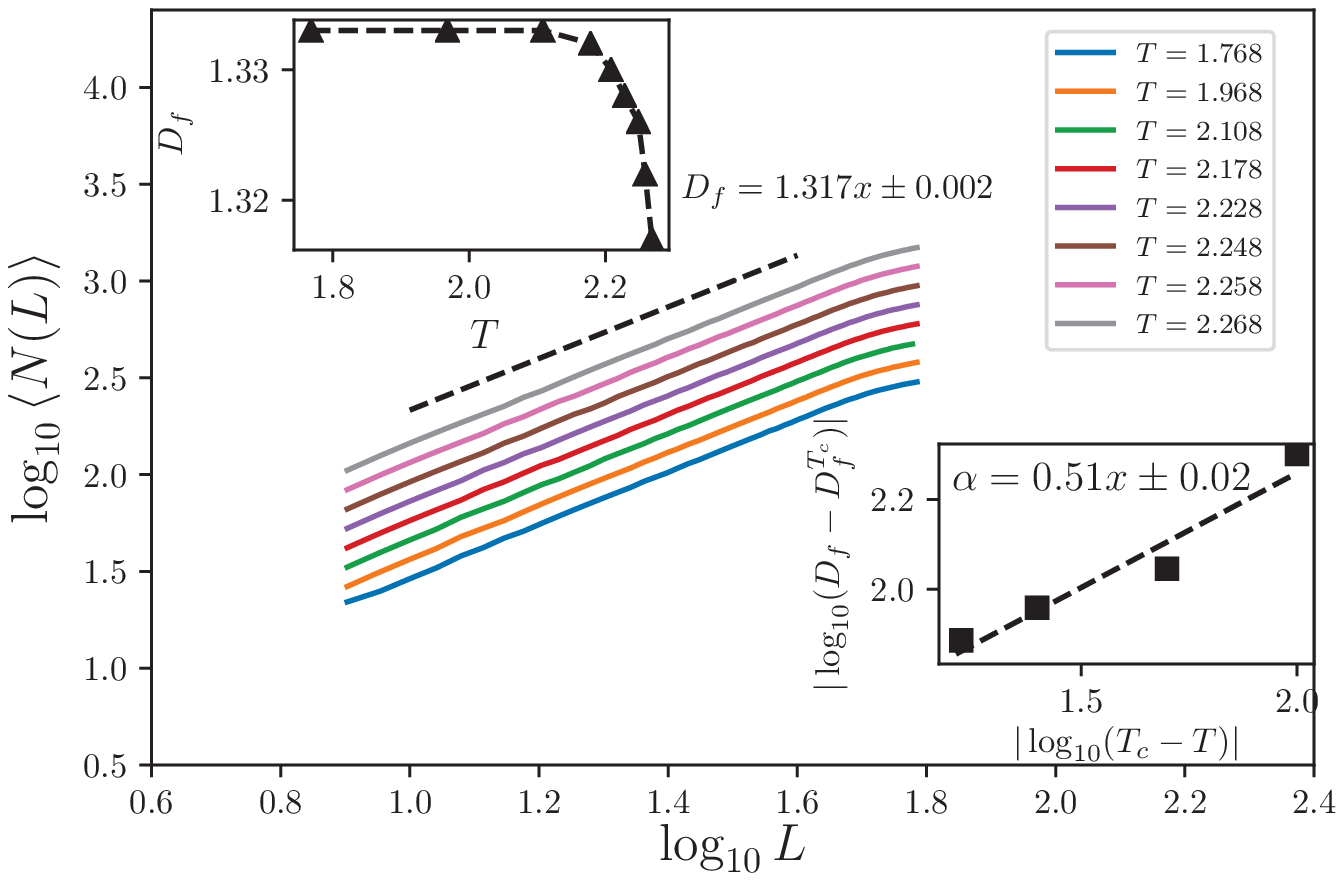}
		\caption{}
		\label{fig:BC}
	\end{subfigure}
	\begin{subfigure}{0.49\textwidth}\includegraphics[width=\textwidth]{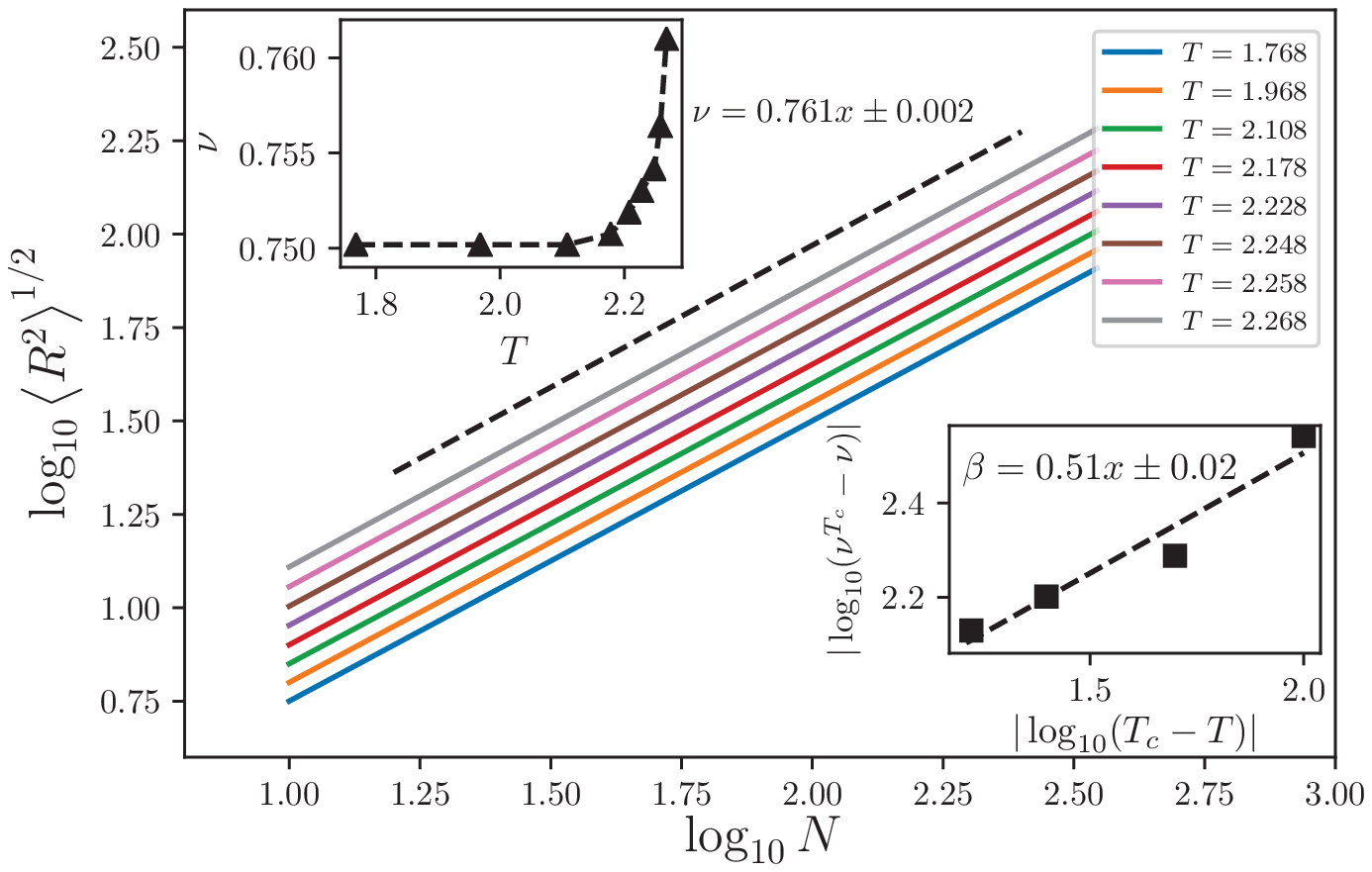}
		\caption{}
		\label{fig:rms}
	\end{subfigure}
	\begin{subfigure}{0.49\textwidth}\includegraphics[width=\textwidth]{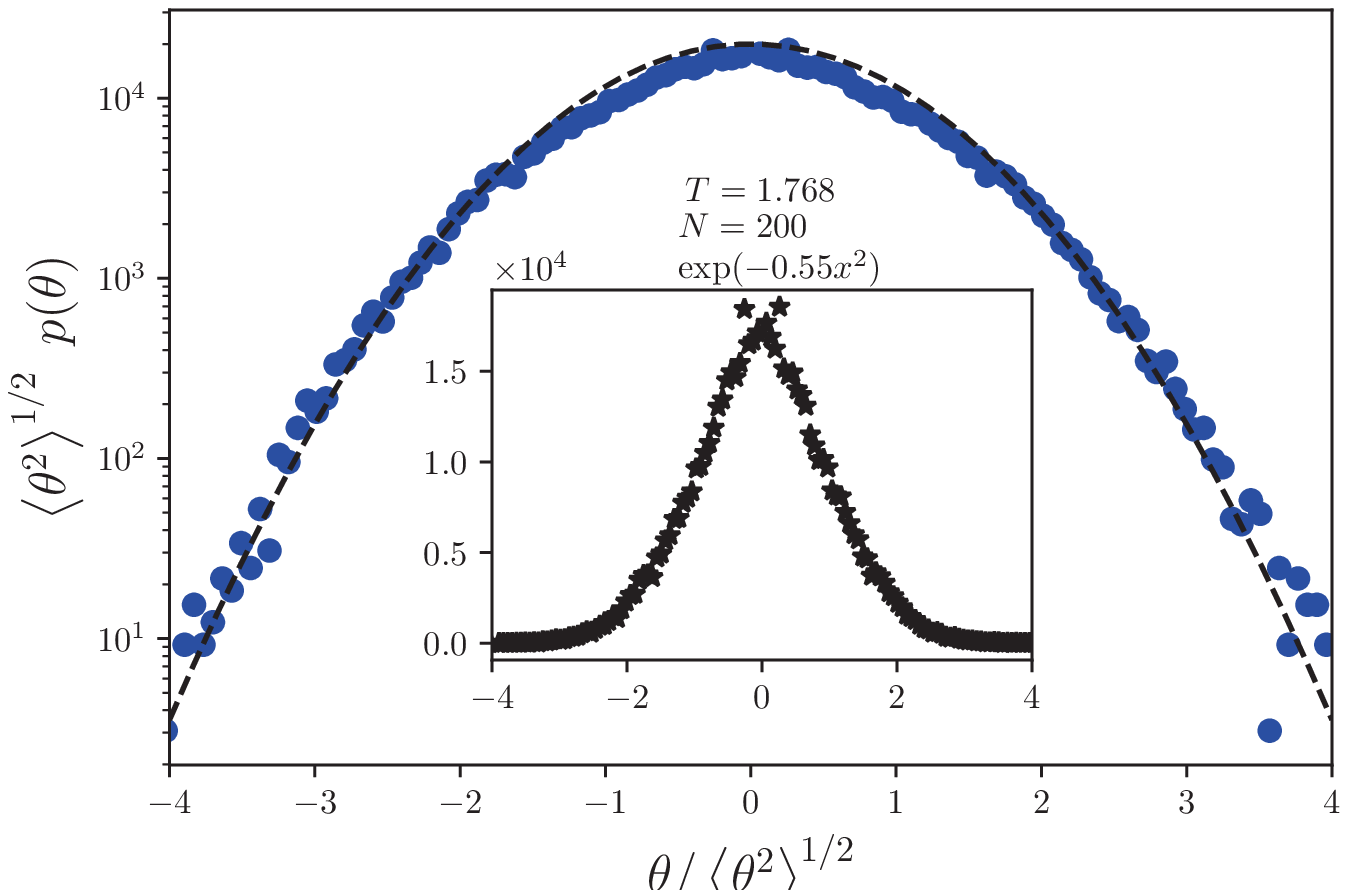}
		\caption{}
		\label{fig:theta}
	\end{subfigure}
	\begin{subfigure}{0.49\textwidth}\includegraphics[width=\textwidth]{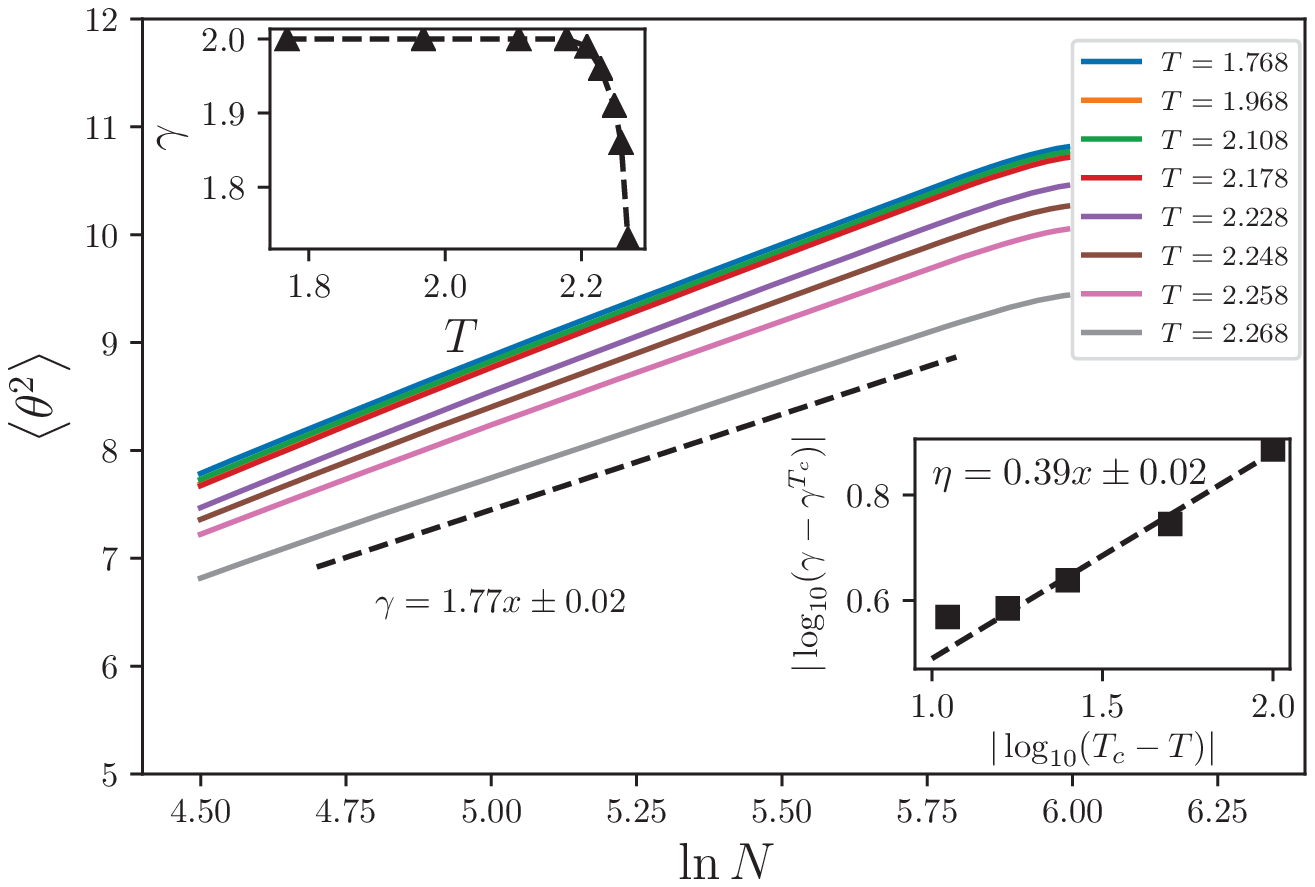}
		\caption{}
		\label{fig:WA}
	\end{subfigure}
	\caption{(Color online) (a) $\log_{10}\left\langle N(L)\right\rangle$ in terms of $\log_{10}L$ for various rates of temperature. Upper inset: $D_F$ in terms of $T$, Lower inset: power-law behavior of the fractal dimension. (b) $\log_{10}\left\langle R^2\right\rangle^{\frac{1}{2}}$ in terms of $\log_{10}N$ for various rates of temperature. Upper inset: the $\nu$ exponent in terms of $T$, Lower inset: power law behavior of $\nu$. (c) The distribution of the winding angel $\theta$ for $N=200$ and $T=1.768$. (d) $\left\langle \theta^2\right\rangle $ in terms of $\ln N$ with the slope $\gamma$ which is $T$-dependent. Upper inset: $\gamma$ in terms of $T$, Lower inset: The power-law behavior of $\gamma$.}
	\label{fig:Off-Tc}
\end{figure*}

The behavior of SAW for all temperatures shows its overall structure. For all temperatures in range the SAWs show power-law behaviors with some well-defined exponents. Note that $T=0$ is the regular system. Our observations show that the exponents are rapidly saturated and become nearly constant with small fluctuations for low-temperatures towards the $T=0$ results. For example Fig.~\ref{fig:BC} shows this behavior (see the upper inset) for the fractal dimension obtained by box-counting method ($D_F(T)$). Interestingly we have observed that in the vicinity of the critical temperature some power-law behaviors arise in terms of $|T-T_c|$. The lower inset of this figure reveals that:
\begin{equation}
\log_{10}|D_F(T)-D_F(T_c)|=\alpha \log_{10}\epsilon+\text{const.}
\end{equation} 
in which $\epsilon\equiv \frac{|T-T_c|}{T_c}$ and $\alpha$ is a new exponent whose value has been obtained using the least squares estimator (LSE) method. The same feature is seen for $\nu(T)$ exponent from the analyzing of $\sqrt{\left\langle R^2\right\rangle} $ (Fig.~\ref{fig:rms}), in which the corresponding exponent $\beta$ is the same as $\alpha= 0.51\pm 0.02$ that is understood by the relation $D_F(T)=\frac{1}{\nu(T)}$. \\
Now let us consider the winding angel statistics. This quantity shows a Gaussian distribution with variance proportional to logarithm of the length of chain. As an example we have shown $\left\langle \theta^2\right\rangle^{\frac{1}{2}}p(\theta)$ in terms of $\theta/\left\langle \theta^2\right\rangle^{\frac{1}{2}}$ for $N=200$ in Fig.~\ref{fig:theta} which is apparently Gaussian. We have calculated the slope of $\left\langle \theta^2\right\rangle $ in terms of $\ln N$ (the $\gamma$ exponent in Fig.~\ref{fig:WA}) which is $2$ for low temperatures. In the vicinity of $T_c$ one can easily show that:
\begin{equation}
\gamma(T)-\gamma(T_c)=A\epsilon^{\alpha}+B\epsilon^{2\alpha}
\end{equation} 
For the temperatures very close to $T_c$ the first term of right hand side is dominant, which leads to a power-law behavior (lower inset of Fig.~\ref{fig:WA}). Therefore the exponent for the temperatures close to $T_c$ should be more or less equal to $\alpha$ as is seen. Note that the discrepancy comes from the non-linear behavior for lower temperatures.\\

The main finding of this section is therefore a new exponent $\alpha$ whose closest fractional value is $\frac{1}{2}$. Noting that the correlation length of the 2D Ising model $\xi$ scales with temperature as $\xi\sim |T-T_c|^{-1}$, one finds the following scaling relation:
\begin{equation}
D_F^{\text{SAW}}(T)-D_F^{\text{SAW}}(T_c)\sim \frac{1}{\sqrt{\xi}}.
\end{equation}

\section*{Discussion and Conclusion}
\label{sec:conc}

Many features of the random walks on the random fractal lattices are known. Since these host systems have commonly considered to be uncorrelated, introducing correlation in the host system is important and interesting which leads to some non-trivial effects on the statistics of self-avoiding random walkers. In this paper we have considered the SAW on the Ising-correlated site-diluted percolation lattice whose correlations are controlled by the temperature $T$. The importance of the correlations has been argued in terms of a stochastic differential equation. The enriched Rosenbluth method as well as winding angel statistics have been employed to obtain the critical exponents of the system in both $T=T_c$ and $T<T_c$ cases. We found that the exponents at $T=T_c$ are in agreement with the Flory's approximation. The winding angel analysis more or less showed the same features. This suggests that the $\nu$ exponent depends only on the effective fractal dimension of the host system. For the temperatures in the range $T<T_c$ some other interesting power-law behaviors have arisen. The calculated exponents reveals that $D_F(T)-D_F(T_c)\sim \frac{1}{\sqrt{\xi(T)}}$ in which $\xi(T)$ is the correlation length of the off-critical Ising system. The winding angel statistics also confirms this result.

\bibliography{refs}

\begin{thebibliography}{47}%
\makeatletter
\providecommand \@ifxundefined [1]{%
 \@ifx{#1\undefined}
}%
\providecommand \@ifnum [1]{%
 \ifnum #1\expandafter \@firstoftwo
 \else \expandafter \@secondoftwo
 \fi
}%
\providecommand \@ifx [1]{%
 \ifx #1\expandafter \@firstoftwo
 \else \expandafter \@secondoftwo
 \fi
}%
\providecommand \natexlab [1]{#1}%
\providecommand \enquote  [1]{``#1''}%
\providecommand \bibnamefont  [1]{#1}%
\providecommand \bibfnamefont [1]{#1}%
\providecommand \citenamefont [1]{#1}%
\providecommand \href@noop [0]{\@secondoftwo}%
\providecommand \href [0]{\begingroup \@sanitize@url \@href}%
\providecommand \@href[1]{\@@startlink{#1}\@@href}%
\providecommand \@@href[1]{\endgroup#1\@@endlink}%
\providecommand \@sanitize@url [0]{\catcode `\\12\catcode `\$12\catcode
  `\&12\catcode `\#12\catcode `\^12\catcode `\_12\catcode `\%12\relax}%
\providecommand \@@startlink[1]{}%
\providecommand \@@endlink[0]{}%
\providecommand \url  [0]{\begingroup\@sanitize@url \@url }%
\providecommand \@url [1]{\endgroup\@href {#1}{\urlprefix }}%
\providecommand \urlprefix  [0]{URL }%
\providecommand \Eprint [0]{\href }%
\providecommand \doibase [0]{http://dx.doi.org/}%
\providecommand \selectlanguage [0]{\@gobble}%
\providecommand \bibinfo  [0]{\@secondoftwo}%
\providecommand \bibfield  [0]{\@secondoftwo}%
\providecommand \translation [1]{[#1]}%
\providecommand \BibitemOpen [0]{}%
\providecommand \bibitemStop [0]{}%
\providecommand \bibitemNoStop [0]{.\EOS\space}%
\providecommand \EOS [0]{\spacefactor3000\relax}%
\providecommand \BibitemShut  [1]{\csname bibitem#1\endcsname}%
\let\auto@bib@innerbib\@empty
\bibitem [{\citenamefont {Barat}\ and\ \citenamefont
  {Chakrabarti}(1995)}]{barat1995statistics}%
  \BibitemOpen
  \bibfield  {author} {\bibinfo {author} {\bibfnamefont {K.}~\bibnamefont
  {Barat}}\ and\ \bibinfo {author} {\bibfnamefont {B.~K.}\ \bibnamefont
  {Chakrabarti}},\ }\href@noop {} {\bibfield  {journal} {\bibinfo  {journal}
  {Physics Reports}\ }\textbf {\bibinfo {volume} {258}},\ \bibinfo {pages}
  {377} (\bibinfo {year} {1995})}\BibitemShut {NoStop}%
\bibitem [{\citenamefont {Rammal}\ \emph {et~al.}(1984)\citenamefont {Rammal},
  \citenamefont {Toulouse},\ and\ \citenamefont {Vannimenus}}]{rammal1984self}%
  \BibitemOpen
  \bibfield  {author} {\bibinfo {author} {\bibfnamefont {R.}~\bibnamefont
  {Rammal}}, \bibinfo {author} {\bibfnamefont {G.}~\bibnamefont {Toulouse}}, \
  and\ \bibinfo {author} {\bibfnamefont {J.}~\bibnamefont {Vannimenus}},\
  }\href@noop {} {\bibfield  {journal} {\bibinfo  {journal} {Journal de
  Physique}\ }\textbf {\bibinfo {volume} {45}},\ \bibinfo {pages} {389}
  (\bibinfo {year} {1984})}\BibitemShut {NoStop}%
\bibitem [{\citenamefont {Chakrabarti}\ and\ \citenamefont
  {Kertesz}(1981)}]{chakrabarti1981statistics}%
  \BibitemOpen
  \bibfield  {author} {\bibinfo {author} {\bibfnamefont {B.}~\bibnamefont
  {Chakrabarti}}\ and\ \bibinfo {author} {\bibfnamefont {J.}~\bibnamefont
  {Kertesz}},\ }\href@noop {} {\bibfield  {journal} {\bibinfo  {journal}
  {Zeitschrift f{\"u}r Physik B Condensed Matter}\ }\textbf {\bibinfo {volume}
  {44}},\ \bibinfo {pages} {221} (\bibinfo {year} {1981})}\BibitemShut
  {NoStop}%
\bibitem [{\citenamefont {Kremer}(1981)}]{kremer1981self}%
  \BibitemOpen
  \bibfield  {author} {\bibinfo {author} {\bibfnamefont {K.}~\bibnamefont
  {Kremer}},\ }\href@noop {} {\bibfield  {journal} {\bibinfo  {journal}
  {Zeitschrift f{\"u}r Physik B Condensed Matter}\ }\textbf {\bibinfo {volume}
  {45}},\ \bibinfo {pages} {149} (\bibinfo {year} {1981})}\BibitemShut
  {NoStop}%
\bibitem [{\citenamefont {Aharony}\ and\ \citenamefont
  {Harris}(1989)}]{aharony1989flory}%
  \BibitemOpen
  \bibfield  {author} {\bibinfo {author} {\bibfnamefont {A.}~\bibnamefont
  {Aharony}}\ and\ \bibinfo {author} {\bibfnamefont {A.~B.}\ \bibnamefont
  {Harris}},\ }\href@noop {} {\bibfield  {journal} {\bibinfo  {journal}
  {Journal of Statistical Physics}\ }\textbf {\bibinfo {volume} {54}},\
  \bibinfo {pages} {1091} (\bibinfo {year} {1989})}\BibitemShut {NoStop}%
\bibitem [{\citenamefont {Elezovic}\ \emph {et~al.}(1987)\citenamefont
  {Elezovic}, \citenamefont {Knezevic},\ and\ \citenamefont
  {Milosevic}}]{elezovic1987critical}%
  \BibitemOpen
  \bibfield  {author} {\bibinfo {author} {\bibfnamefont {S.}~\bibnamefont
  {Elezovic}}, \bibinfo {author} {\bibfnamefont {M.}~\bibnamefont {Knezevic}},
  \ and\ \bibinfo {author} {\bibfnamefont {S.}~\bibnamefont {Milosevic}},\
  }\href@noop {} {\bibfield  {journal} {\bibinfo  {journal} {Journal of Physics
  A: Mathematical and General}\ }\textbf {\bibinfo {volume} {20}},\ \bibinfo
  {pages} {1215} (\bibinfo {year} {1987})}\BibitemShut {NoStop}%
\bibitem [{\citenamefont {Dhar}(1988)}]{dhar1988critical}%
  \BibitemOpen
  \bibfield  {author} {\bibinfo {author} {\bibfnamefont {D.}~\bibnamefont
  {Dhar}},\ }\href@noop {} {\bibfield  {journal} {\bibinfo  {journal} {Journal
  de Physique}\ }\textbf {\bibinfo {volume} {49}},\ \bibinfo {pages} {397}
  (\bibinfo {year} {1988})}\BibitemShut {NoStop}%
\bibitem [{\citenamefont {Milosevic}\ and\ \citenamefont
  {Zivic}(1991)}]{milosevic1991self}%
  \BibitemOpen
  \bibfield  {author} {\bibinfo {author} {\bibfnamefont {S.}~\bibnamefont
  {Milosevic}}\ and\ \bibinfo {author} {\bibfnamefont {I.}~\bibnamefont
  {Zivic}},\ }\href@noop {} {\bibfield  {journal} {\bibinfo  {journal} {Journal
  of Physics A: Mathematical and General}\ }\textbf {\bibinfo {volume} {24}},\
  \bibinfo {pages} {L833} (\bibinfo {year} {1991})}\BibitemShut {NoStop}%
\bibitem [{\citenamefont {Lam}\ and\ \citenamefont
  {Zhang}(1984)}]{lam1984self}%
  \BibitemOpen
  \bibfield  {author} {\bibinfo {author} {\bibfnamefont {P.}~\bibnamefont
  {Lam}}\ and\ \bibinfo {author} {\bibfnamefont {Z.}~\bibnamefont {Zhang}},\
  }\href@noop {} {\bibfield  {journal} {\bibinfo  {journal} {Zeitschrift
  f{\"u}r Physik B Condensed Matter}\ }\textbf {\bibinfo {volume} {56}},\
  \bibinfo {pages} {155} (\bibinfo {year} {1984})}\BibitemShut {NoStop}%
\bibitem [{\citenamefont {Roy}\ and\ \citenamefont
  {Chakrabarti}(1987)}]{roy1987scaling}%
  \BibitemOpen
  \bibfield  {author} {\bibinfo {author} {\bibfnamefont {A.}~\bibnamefont
  {Roy}}\ and\ \bibinfo {author} {\bibfnamefont {B.}~\bibnamefont
  {Chakrabarti}},\ }\href@noop {} {\bibfield  {journal} {\bibinfo  {journal}
  {Journal of Physics A: Mathematical and General}\ }\textbf {\bibinfo {volume}
  {20}},\ \bibinfo {pages} {215} (\bibinfo {year} {1987})}\BibitemShut
  {NoStop}%
\bibitem [{\citenamefont {Roy}\ and\ \citenamefont
  {Blumen}(1990)}]{roy1990theory}%
  \BibitemOpen
  \bibfield  {author} {\bibinfo {author} {\bibfnamefont {A.}~\bibnamefont
  {Roy}}\ and\ \bibinfo {author} {\bibfnamefont {A.}~\bibnamefont {Blumen}},\
  }\href@noop {} {\bibfield  {journal} {\bibinfo  {journal} {Journal of
  Statistical Physics}\ }\textbf {\bibinfo {volume} {59}},\ \bibinfo {pages}
  {1581} (\bibinfo {year} {1990})}\BibitemShut {NoStop}%
\bibitem [{\citenamefont {Blavatska}\ and\ \citenamefont
  {Janke}(2008{\natexlab{a}})}]{blavatska2008scaling}%
  \BibitemOpen
  \bibfield  {author} {\bibinfo {author} {\bibfnamefont {V.}~\bibnamefont
  {Blavatska}}\ and\ \bibinfo {author} {\bibfnamefont {W.}~\bibnamefont
  {Janke}},\ }\href@noop {} {\bibfield  {journal} {\bibinfo  {journal} {EPL
  (Europhysics Letters)}\ }\textbf {\bibinfo {volume} {82}},\ \bibinfo {pages}
  {66006} (\bibinfo {year} {2008}{\natexlab{a}})}\BibitemShut {NoStop}%
\bibitem [{\citenamefont {Blavatska}\ and\ \citenamefont
  {Janke}(2008{\natexlab{b}})}]{blavatska2008walking}%
  \BibitemOpen
  \bibfield  {author} {\bibinfo {author} {\bibfnamefont {V.}~\bibnamefont
  {Blavatska}}\ and\ \bibinfo {author} {\bibfnamefont {W.}~\bibnamefont
  {Janke}},\ }\href@noop {} {\bibfield  {journal} {\bibinfo  {journal} {Journal
  of Physics A: Mathematical and Theoretical}\ }\textbf {\bibinfo {volume}
  {42}},\ \bibinfo {pages} {015001} (\bibinfo {year}
  {2008}{\natexlab{b}})}\BibitemShut {NoStop}%
\bibitem [{\citenamefont {Lee}\ and\ \citenamefont
  {Nakanishi}(1988)}]{lee1988self}%
  \BibitemOpen
  \bibfield  {author} {\bibinfo {author} {\bibfnamefont {S.~B.}\ \bibnamefont
  {Lee}}\ and\ \bibinfo {author} {\bibfnamefont {H.}~\bibnamefont
  {Nakanishi}},\ }\href@noop {} {\bibfield  {journal} {\bibinfo  {journal}
  {Physical review letters}\ }\textbf {\bibinfo {volume} {61}},\ \bibinfo
  {pages} {2022} (\bibinfo {year} {1988})}\BibitemShut {NoStop}%
\bibitem [{\citenamefont {Lee}\ \emph {et~al.}(1989)\citenamefont {Lee},
  \citenamefont {Nakanishi},\ and\ \citenamefont {Kim}}]{lee1989monte}%
  \BibitemOpen
  \bibfield  {author} {\bibinfo {author} {\bibfnamefont {S.~B.}\ \bibnamefont
  {Lee}}, \bibinfo {author} {\bibfnamefont {H.}~\bibnamefont {Nakanishi}}, \
  and\ \bibinfo {author} {\bibfnamefont {Y.}~\bibnamefont {Kim}},\ }\href@noop
  {} {\bibfield  {journal} {\bibinfo  {journal} {Physical Review B}\ }\textbf
  {\bibinfo {volume} {39}},\ \bibinfo {pages} {9561} (\bibinfo {year}
  {1989})}\BibitemShut {NoStop}%
\bibitem [{\citenamefont {Nakanishi}\ and\ \citenamefont
  {Moon}(1992)}]{nakanishi1992self}%
  \BibitemOpen
  \bibfield  {author} {\bibinfo {author} {\bibfnamefont {H.}~\bibnamefont
  {Nakanishi}}\ and\ \bibinfo {author} {\bibfnamefont {J.}~\bibnamefont
  {Moon}},\ }\href@noop {} {\bibfield  {journal} {\bibinfo  {journal} {Physica
  A: Statistical Mechanics and its Applications}\ }\textbf {\bibinfo {volume}
  {191}},\ \bibinfo {pages} {309} (\bibinfo {year} {1992})}\BibitemShut
  {NoStop}%
\bibitem [{\citenamefont {Rintoul}\ \emph {et~al.}(1994)\citenamefont
  {Rintoul}, \citenamefont {Moon},\ and\ \citenamefont
  {Nakanishi}}]{rintoul1994statistics}%
  \BibitemOpen
  \bibfield  {author} {\bibinfo {author} {\bibfnamefont {M.}~\bibnamefont
  {Rintoul}}, \bibinfo {author} {\bibfnamefont {J.}~\bibnamefont {Moon}}, \
  and\ \bibinfo {author} {\bibfnamefont {H.}~\bibnamefont {Nakanishi}},\
  }\href@noop {} {\bibfield  {journal} {\bibinfo  {journal} {Physical Review
  E}\ }\textbf {\bibinfo {volume} {49}},\ \bibinfo {pages} {2790} (\bibinfo
  {year} {1994})}\BibitemShut {NoStop}%
\bibitem [{\citenamefont {Meir}\ and\ \citenamefont
  {Harris}(1989)}]{meir1989self}%
  \BibitemOpen
  \bibfield  {author} {\bibinfo {author} {\bibfnamefont {Y.}~\bibnamefont
  {Meir}}\ and\ \bibinfo {author} {\bibfnamefont {A.~B.}\ \bibnamefont
  {Harris}},\ }\href@noop {} {\bibfield  {journal} {\bibinfo  {journal}
  {Physical review letters}\ }\textbf {\bibinfo {volume} {63}},\ \bibinfo
  {pages} {2819} (\bibinfo {year} {1989})}\BibitemShut {NoStop}%
\bibitem [{\citenamefont {Von~Ferber}\ \emph {et~al.}(2004)\citenamefont
  {Von~Ferber}, \citenamefont {Blavats’ka}, \citenamefont {Folk},\ and\
  \citenamefont {Holovatch}}]{von2004two}%
  \BibitemOpen
  \bibfield  {author} {\bibinfo {author} {\bibfnamefont {C.}~\bibnamefont
  {Von~Ferber}}, \bibinfo {author} {\bibfnamefont {V.}~\bibnamefont
  {Blavats’ka}}, \bibinfo {author} {\bibfnamefont {R.}~\bibnamefont {Folk}},
  \ and\ \bibinfo {author} {\bibfnamefont {Y.}~\bibnamefont {Holovatch}},\
  }\href@noop {} {\bibfield  {journal} {\bibinfo  {journal} {Physical Review
  E}\ }\textbf {\bibinfo {volume} {70}},\ \bibinfo {pages} {035104} (\bibinfo
  {year} {2004})}\BibitemShut {NoStop}%
\bibitem [{\citenamefont {Sahimi}(1984)}]{sahimi1984self}%
  \BibitemOpen
  \bibfield  {author} {\bibinfo {author} {\bibfnamefont {M.}~\bibnamefont
  {Sahimi}},\ }\href@noop {} {\bibfield  {journal} {\bibinfo  {journal}
  {Journal of Physics A: Mathematical and General}\ }\textbf {\bibinfo {volume}
  {17}},\ \bibinfo {pages} {L379} (\bibinfo {year} {1984})}\BibitemShut
  {NoStop}%
\bibitem [{\citenamefont {Gefen}\ \emph {et~al.}(1980)\citenamefont {Gefen},
  \citenamefont {Mandelbrot},\ and\ \citenamefont
  {Aharony}}]{gefen1980critical}%
  \BibitemOpen
  \bibfield  {author} {\bibinfo {author} {\bibfnamefont {Y.}~\bibnamefont
  {Gefen}}, \bibinfo {author} {\bibfnamefont {B.~B.}\ \bibnamefont
  {Mandelbrot}}, \ and\ \bibinfo {author} {\bibfnamefont {A.}~\bibnamefont
  {Aharony}},\ }\href@noop {} {\bibfield  {journal} {\bibinfo  {journal}
  {Physical Review Letters}\ }\textbf {\bibinfo {volume} {45}},\ \bibinfo
  {pages} {855} (\bibinfo {year} {1980})}\BibitemShut {NoStop}%
\bibitem [{\citenamefont {Kose}\ \emph {et~al.}(2009)\citenamefont {Kose},
  \citenamefont {Fischer}, \citenamefont {Mao},\ and\ \citenamefont
  {Koser}}]{kose2009label}%
  \BibitemOpen
  \bibfield  {author} {\bibinfo {author} {\bibfnamefont {A.~R.}\ \bibnamefont
  {Kose}}, \bibinfo {author} {\bibfnamefont {B.}~\bibnamefont {Fischer}},
  \bibinfo {author} {\bibfnamefont {L.}~\bibnamefont {Mao}}, \ and\ \bibinfo
  {author} {\bibfnamefont {H.}~\bibnamefont {Koser}},\ }\href@noop {}
  {\bibfield  {journal} {\bibinfo  {journal} {Proceedings of the National
  Academy of Sciences}\ }\textbf {\bibinfo {volume} {106}},\ \bibinfo {pages}
  {21478} (\bibinfo {year} {2009})}\BibitemShut {NoStop}%
\bibitem [{\citenamefont {Kikura}\ \emph {et~al.}(2004)\citenamefont {Kikura},
  \citenamefont {Matsushita}, \citenamefont {Matsuzaki}, \citenamefont
  {Kobayashi},\ and\ \citenamefont {Aritomi}}]{kikura2004thermal}%
  \BibitemOpen
  \bibfield  {author} {\bibinfo {author} {\bibfnamefont {H.}~\bibnamefont
  {Kikura}}, \bibinfo {author} {\bibfnamefont {J.}~\bibnamefont {Matsushita}},
  \bibinfo {author} {\bibfnamefont {M.}~\bibnamefont {Matsuzaki}}, \bibinfo
  {author} {\bibfnamefont {Y.}~\bibnamefont {Kobayashi}}, \ and\ \bibinfo
  {author} {\bibfnamefont {M.}~\bibnamefont {Aritomi}},\ }\href@noop {}
  {\bibfield  {journal} {\bibinfo  {journal} {Science and Technology of
  Advanced Materials}\ }\textbf {\bibinfo {volume} {5}},\ \bibinfo {pages}
  {703} (\bibinfo {year} {2004})}\BibitemShut {NoStop}%
\bibitem [{\citenamefont {Matsuzaki}\ \emph {et~al.}(2004)\citenamefont
  {Matsuzaki}, \citenamefont {Kikura}, \citenamefont {Matsushita},
  \citenamefont {Aritomi},\ and\ \citenamefont {Akatsuka}}]{matsuzaki2004real}%
  \BibitemOpen
  \bibfield  {author} {\bibinfo {author} {\bibfnamefont {M.}~\bibnamefont
  {Matsuzaki}}, \bibinfo {author} {\bibfnamefont {H.}~\bibnamefont {Kikura}},
  \bibinfo {author} {\bibfnamefont {J.}~\bibnamefont {Matsushita}}, \bibinfo
  {author} {\bibfnamefont {M.}~\bibnamefont {Aritomi}}, \ and\ \bibinfo
  {author} {\bibfnamefont {H.}~\bibnamefont {Akatsuka}},\ }\href@noop {}
  {\bibfield  {journal} {\bibinfo  {journal} {Science and Technology of
  Advanced Materials}\ }\textbf {\bibinfo {volume} {5}},\ \bibinfo {pages}
  {667} (\bibinfo {year} {2004})}\BibitemShut {NoStop}%
\bibitem [{\citenamefont {Philip}\ \emph {et~al.}(2007)\citenamefont {Philip},
  \citenamefont {Shima},\ and\ \citenamefont {Raj}}]{philip2007enhancement}%
  \BibitemOpen
  \bibfield  {author} {\bibinfo {author} {\bibfnamefont {J.}~\bibnamefont
  {Philip}}, \bibinfo {author} {\bibfnamefont {P.}~\bibnamefont {Shima}}, \
  and\ \bibinfo {author} {\bibfnamefont {B.}~\bibnamefont {Raj}},\ }\href@noop
  {} {\bibfield  {journal} {\bibinfo  {journal} {Applied physics letters}\
  }\textbf {\bibinfo {volume} {91}},\ \bibinfo {pages} {203108} (\bibinfo
  {year} {2007})}\BibitemShut {NoStop}%
\bibitem [{\citenamefont {Kim}\ \emph {et~al.}(2008)\citenamefont {Kim},
  \citenamefont {Fang}, \citenamefont {Choi},\ and\ \citenamefont
  {Seo}}]{kim2008magnetic}%
  \BibitemOpen
  \bibfield  {author} {\bibinfo {author} {\bibfnamefont {J.~H.}\ \bibnamefont
  {Kim}}, \bibinfo {author} {\bibfnamefont {F.~F.}\ \bibnamefont {Fang}},
  \bibinfo {author} {\bibfnamefont {H.~J.}\ \bibnamefont {Choi}}, \ and\
  \bibinfo {author} {\bibfnamefont {Y.}~\bibnamefont {Seo}},\ }\href@noop {}
  {\bibfield  {journal} {\bibinfo  {journal} {Materials Letters}\ }\textbf
  {\bibinfo {volume} {62}},\ \bibinfo {pages} {2897} (\bibinfo {year}
  {2008})}\BibitemShut {NoStop}%
\bibitem [{\citenamefont {Keng}\ \emph {et~al.}(2009)\citenamefont {Keng},
  \citenamefont {Kim}, \citenamefont {Shim}, \citenamefont {Sahoo},
  \citenamefont {Veneman}, \citenamefont {Armstrong}, \citenamefont {Yoo},
  \citenamefont {Pemberton}, \citenamefont {Bull}, \citenamefont {Griebel}
  \emph {et~al.}}]{keng2009colloidal}%
  \BibitemOpen
  \bibfield  {author} {\bibinfo {author} {\bibfnamefont {P.~Y.}\ \bibnamefont
  {Keng}}, \bibinfo {author} {\bibfnamefont {B.~Y.}\ \bibnamefont {Kim}},
  \bibinfo {author} {\bibfnamefont {I.-B.}\ \bibnamefont {Shim}}, \bibinfo
  {author} {\bibfnamefont {R.}~\bibnamefont {Sahoo}}, \bibinfo {author}
  {\bibfnamefont {P.~E.}\ \bibnamefont {Veneman}}, \bibinfo {author}
  {\bibfnamefont {N.~R.}\ \bibnamefont {Armstrong}}, \bibinfo {author}
  {\bibfnamefont {H.}~\bibnamefont {Yoo}}, \bibinfo {author} {\bibfnamefont
  {J.~E.}\ \bibnamefont {Pemberton}}, \bibinfo {author} {\bibfnamefont {M.~M.}\
  \bibnamefont {Bull}}, \bibinfo {author} {\bibfnamefont {J.~J.}\ \bibnamefont
  {Griebel}},  \emph {et~al.},\ }\href@noop {} {\bibfield  {journal} {\bibinfo
  {journal} {ACS nano}\ }\textbf {\bibinfo {volume} {3}},\ \bibinfo {pages}
  {3143} (\bibinfo {year} {2009})}\BibitemShut {NoStop}%
\bibitem [{\citenamefont {Kikura}\ \emph {et~al.}(2007)\citenamefont {Kikura},
  \citenamefont {Matsushita}, \citenamefont {Kakuta}, \citenamefont {Aritomi},\
  and\ \citenamefont {Kobayashi}}]{kikura2007cluster}%
  \BibitemOpen
  \bibfield  {author} {\bibinfo {author} {\bibfnamefont {H.}~\bibnamefont
  {Kikura}}, \bibinfo {author} {\bibfnamefont {J.}~\bibnamefont {Matsushita}},
  \bibinfo {author} {\bibfnamefont {N.}~\bibnamefont {Kakuta}}, \bibinfo
  {author} {\bibfnamefont {M.}~\bibnamefont {Aritomi}}, \ and\ \bibinfo
  {author} {\bibfnamefont {Y.}~\bibnamefont {Kobayashi}},\ }\href@noop {}
  {\bibfield  {journal} {\bibinfo  {journal} {Journal of materials processing
  technology}\ }\textbf {\bibinfo {volume} {181}},\ \bibinfo {pages} {93}
  (\bibinfo {year} {2007})}\BibitemShut {NoStop}%
\bibitem [{\citenamefont {Najafi}(2016{\natexlab{a}})}]{najafi2016monte}%
  \BibitemOpen
  \bibfield  {author} {\bibinfo {author} {\bibfnamefont {M.}~\bibnamefont
  {Najafi}},\ }\href@noop {} {\bibfield  {journal} {\bibinfo  {journal}
  {Physics Letters A}\ }\textbf {\bibinfo {volume} {380}},\ \bibinfo {pages}
  {370} (\bibinfo {year} {2016}{\natexlab{a}})}\BibitemShut {NoStop}%
\bibitem [{\citenamefont {Najafi}\ \emph {et~al.}(2016)\citenamefont {Najafi},
  \citenamefont {Ghaedi},\ and\ \citenamefont
  {Moghimi-Araghi}}]{najafi2016water}%
  \BibitemOpen
  \bibfield  {author} {\bibinfo {author} {\bibfnamefont {M.}~\bibnamefont
  {Najafi}}, \bibinfo {author} {\bibfnamefont {M.}~\bibnamefont {Ghaedi}}, \
  and\ \bibinfo {author} {\bibfnamefont {S.}~\bibnamefont {Moghimi-Araghi}},\
  }\href@noop {} {\bibfield  {journal} {\bibinfo  {journal} {Physica A:
  Statistical Mechanics and its Applications}\ }\textbf {\bibinfo {volume}
  {445}},\ \bibinfo {pages} {102} (\bibinfo {year} {2016})}\BibitemShut
  {NoStop}%
\bibitem [{\citenamefont {Najafi}\ and\ \citenamefont
  {Ghaedi}(2015)}]{najafi2015geometrical}%
  \BibitemOpen
  \bibfield  {author} {\bibinfo {author} {\bibfnamefont {M.}~\bibnamefont
  {Najafi}}\ and\ \bibinfo {author} {\bibfnamefont {M.}~\bibnamefont
  {Ghaedi}},\ }\href@noop {} {\bibfield  {journal} {\bibinfo  {journal}
  {Physica A: Statistical Mechanics and its Applications}\ }\textbf {\bibinfo
  {volume} {427}},\ \bibinfo {pages} {82} (\bibinfo {year} {2015})}\BibitemShut
  {NoStop}%
\bibitem [{\citenamefont {Najafi}(2016{\natexlab{b}})}]{najafi2016bak}%
  \BibitemOpen
  \bibfield  {author} {\bibinfo {author} {\bibfnamefont {M.}~\bibnamefont
  {Najafi}},\ }\href@noop {} {\bibfield  {journal} {\bibinfo  {journal}
  {Journal of Physics A: Mathematical and Theoretical}\ }\textbf {\bibinfo
  {volume} {49}},\ \bibinfo {pages} {335003} (\bibinfo {year}
  {2016}{\natexlab{b}})}\BibitemShut {NoStop}%
\bibitem [{\citenamefont {Cheraghalizadeh}\ \emph {et~al.}(2017)\citenamefont
  {Cheraghalizadeh}, \citenamefont {Najafi}, \citenamefont
  {Dashti-Naserabadi},\ and\ \citenamefont
  {Mohammadzadeh}}]{cheraghalizadeh2017mapping}%
  \BibitemOpen
  \bibfield  {author} {\bibinfo {author} {\bibfnamefont {J.}~\bibnamefont
  {Cheraghalizadeh}}, \bibinfo {author} {\bibfnamefont {M.}~\bibnamefont
  {Najafi}}, \bibinfo {author} {\bibfnamefont {H.}~\bibnamefont
  {Dashti-Naserabadi}}, \ and\ \bibinfo {author} {\bibfnamefont
  {H.}~\bibnamefont {Mohammadzadeh}},\ }\href@noop {} {\bibfield  {journal}
  {\bibinfo  {journal} {Physical Review E}\ }\textbf {\bibinfo {volume} {96}},\
  \bibinfo {pages} {052127} (\bibinfo {year} {2017})}\BibitemShut {NoStop}%
\bibitem [{\citenamefont {Lau}\ and\ \citenamefont
  {Dill}(1989)}]{lau1989lattice}%
  \BibitemOpen
  \bibfield  {author} {\bibinfo {author} {\bibfnamefont {K.~F.}\ \bibnamefont
  {Lau}}\ and\ \bibinfo {author} {\bibfnamefont {K.~A.}\ \bibnamefont {Dill}},\
  }\href@noop {} {\bibfield  {journal} {\bibinfo  {journal} {Macromolecules}\
  }\textbf {\bibinfo {volume} {22}},\ \bibinfo {pages} {3986} (\bibinfo {year}
  {1989})}\BibitemShut {NoStop}%
\bibitem [{\citenamefont {Tang}(2000)}]{tang2000simple}%
  \BibitemOpen
  \bibfield  {author} {\bibinfo {author} {\bibfnamefont {C.}~\bibnamefont
  {Tang}},\ }\href@noop {} {\bibfield  {journal} {\bibinfo  {journal} {Physica
  A: Statistical Mechanics and its Applications}\ }\textbf {\bibinfo {volume}
  {288}},\ \bibinfo {pages} {31} (\bibinfo {year} {2000})}\BibitemShut
  {NoStop}%
\bibitem [{\citenamefont {Delfino}(2009)}]{delfino2009field}%
  \BibitemOpen
  \bibfield  {author} {\bibinfo {author} {\bibfnamefont {G.}~\bibnamefont
  {Delfino}},\ }\href@noop {} {\bibfield  {journal} {\bibinfo  {journal}
  {Nuclear Physics B}\ }\textbf {\bibinfo {volume} {818}},\ \bibinfo {pages}
  {196} (\bibinfo {year} {2009})}\BibitemShut {NoStop}%
\bibitem [{\citenamefont {Cardy}(2005)}]{cardy2005sle}%
  \BibitemOpen
  \bibfield  {author} {\bibinfo {author} {\bibfnamefont {J.}~\bibnamefont
  {Cardy}},\ }\href@noop {} {\bibfield  {journal} {\bibinfo  {journal} {Annals
  of Physics}\ }\textbf {\bibinfo {volume} {318}},\ \bibinfo {pages} {81}
  (\bibinfo {year} {2005})}\BibitemShut {NoStop}%
\bibitem [{\citenamefont {Schramm}(2000)}]{Schramm2000Scaling}%
  \BibitemOpen
  \bibfield  {author} {\bibinfo {author} {\bibfnamefont {O.}~\bibnamefont
  {Schramm}},\ }\href {\doibase 10.1007/BF02803524} {\bibfield  {journal}
  {\bibinfo  {journal} {Israel Journal of Mathematics}\ }\textbf {\bibinfo
  {volume} {118}},\ \bibinfo {pages} {221} (\bibinfo {year}
  {2000})}\BibitemShut {NoStop}%
\bibitem [{\citenamefont {L{\"o}wner}(1923)}]{lowner1923untersuchungen}%
  \BibitemOpen
  \bibfield  {author} {\bibinfo {author} {\bibfnamefont {K.}~\bibnamefont
  {L{\"o}wner}},\ }\href@noop {} {\bibfield  {journal} {\bibinfo  {journal}
  {Mathematische Annalen}\ }\textbf {\bibinfo {volume} {89}},\ \bibinfo {pages}
  {103} (\bibinfo {year} {1923})}\BibitemShut {NoStop}%
\bibitem [{\citenamefont {Smirnov}(2007)}]{Smirnov2007Conformal}%
  \BibitemOpen
  \bibfield  {author} {\bibinfo {author} {\bibfnamefont {S.}~\bibnamefont
  {Smirnov}},\ }\href@noop {} {\bibfield  {journal} {\bibinfo  {journal} {arXiv
  preprint arXiv:0708.0039}\ } (\bibinfo {year} {2007})}\BibitemShut {NoStop}%
\bibitem [{\citenamefont {Najafi}(2013)}]{najafi2013left}%
  \BibitemOpen
  \bibfield  {author} {\bibinfo {author} {\bibfnamefont {M.}~\bibnamefont
  {Najafi}},\ }\href@noop {} {\bibfield  {journal} {\bibinfo  {journal}
  {Physical Review E}\ }\textbf {\bibinfo {volume} {87}},\ \bibinfo {pages}
  {062105} (\bibinfo {year} {2013})}\BibitemShut {NoStop}%
\bibitem [{\citenamefont {Najafi}(2015{\natexlab{a}})}]{Najafi2015Fokker}%
  \BibitemOpen
  \bibfield  {author} {\bibinfo {author} {\bibfnamefont {M.}~\bibnamefont
  {Najafi}},\ }\href {http://arxiv.org/abs/1501.00578} {\  (\bibinfo {year}
  {2015}{\natexlab{a}})},\ \Eprint {http://arxiv.org/abs/1501.00578}
  {arXiv:1501.00578 [cond-mat]} \BibitemShut {NoStop}%
\bibitem [{\citenamefont {Najafi}\ \emph {et~al.}(2012)\citenamefont {Najafi},
  \citenamefont {Moghimi-Araghi},\ and\ \citenamefont
  {Rouhani}}]{Najafi2012Observation}%
  \BibitemOpen
  \bibfield  {author} {\bibinfo {author} {\bibfnamefont {M.}~\bibnamefont
  {Najafi}}, \bibinfo {author} {\bibfnamefont {S.}~\bibnamefont
  {Moghimi-Araghi}}, \ and\ \bibinfo {author} {\bibfnamefont {S.}~\bibnamefont
  {Rouhani}},\ }\href {http://stacks.iop.org/1751-8121/45/i=9/a=095001}
  {\bibfield  {journal} {\bibinfo  {journal} {Journal of Physics A:
  Mathematical and Theoretical}\ }\textbf {\bibinfo {volume} {45}},\ \bibinfo
  {pages} {095001} (\bibinfo {year} {2012})}\BibitemShut {NoStop}%
\bibitem [{\citenamefont {Najafi}(2015{\natexlab{b}})}]{najafi2015observation}%
  \BibitemOpen
  \bibfield  {author} {\bibinfo {author} {\bibfnamefont {M.}~\bibnamefont
  {Najafi}},\ }\href@noop {} {\bibfield  {journal} {\bibinfo  {journal}
  {Journal of Statistical Mechanics: Theory and Experiment}\ }\textbf {\bibinfo
  {volume} {2015}},\ \bibinfo {pages} {P05009} (\bibinfo {year}
  {2015}{\natexlab{b}})}\BibitemShut {NoStop}%
\bibitem [{\citenamefont {Duplantier}\ and\ \citenamefont
  {Saleur}(1988)}]{duplantier1988winding}%
  \BibitemOpen
  \bibfield  {author} {\bibinfo {author} {\bibfnamefont {B.}~\bibnamefont
  {Duplantier}}\ and\ \bibinfo {author} {\bibfnamefont {H.}~\bibnamefont
  {Saleur}},\ }\href@noop {} {\bibfield  {journal} {\bibinfo  {journal}
  {Physical Review Letters}\ }\textbf {\bibinfo {volume} {60}},\ \bibinfo
  {pages} {2343} (\bibinfo {year} {1988})}\BibitemShut {NoStop}%
\bibitem [{\citenamefont {Hoshen}\ and\ \citenamefont
  {Kopelman}(1976)}]{hoshen1976percolation}%
  \BibitemOpen
  \bibfield  {author} {\bibinfo {author} {\bibfnamefont {J.}~\bibnamefont
  {Hoshen}}\ and\ \bibinfo {author} {\bibfnamefont {R.}~\bibnamefont
  {Kopelman}},\ }\href@noop {} {\bibfield  {journal} {\bibinfo  {journal}
  {Physical Review B}\ }\textbf {\bibinfo {volume} {14}},\ \bibinfo {pages}
  {3438} (\bibinfo {year} {1976})}\BibitemShut {NoStop}%
\bibitem [{\citenamefont {Duplantier}\ and\ \citenamefont
  {Saleur}(1989)}]{duplantier1989exact}%
  \BibitemOpen
  \bibfield  {author} {\bibinfo {author} {\bibfnamefont {B.}~\bibnamefont
  {Duplantier}}\ and\ \bibinfo {author} {\bibfnamefont {H.}~\bibnamefont
  {Saleur}},\ }\href@noop {} {\bibfield  {journal} {\bibinfo  {journal}
  {Physical review letters}\ }\textbf {\bibinfo {volume} {63}},\ \bibinfo
  {pages} {2536} (\bibinfo {year} {1989})}\BibitemShut {NoStop}%
\end{thebibliography}%

\end{document}